\documentstyle[prd,aps,epsfig]{revtex}
\begin{document}
\draft
\title{Pomeron in diffractive processes
$\gamma^*(Q^2)p\to\rho^0 p$ and \\
$\gamma^*(Q^2)p\to\gamma^*(Q^2) p$  at large $Q^2$:
the onset of pQCD}
\author{V. V. Anisovich, L. G. Dakhno, D. I. Melikhov,
V. A. Nikonov, and M. G. Ryskin}
\date{\today}
\address{St.Petersburg Nuclear Physics Institute, Gatchina, 188350,
Russia}
\maketitle
\begin{abstract}
We study the reactions
$\gamma^*(Q^2)p\to\rho^0 p$ and
$\gamma^*(Q^2)p\to\gamma^*(Q^2) p$ at large
$Q^2$ and  $W^2/Q^2$ and small momentum transfer, $\kappa^2_\perp$, to
the nucleon  where the pomeron exchange
dominates.  At large $Q^2$ the virtual photon selects a hard $q\bar q$
pair, thus selecting the hard pomeron component (the BFKL pomeron).
The amplitudes for both transverse
and longitudinal polarizations of the initial photon and outgoing
$\rho$-meson (photon) are calculated in the framework of the
BFKL pomeron exchange.
Our calculations show that  one cannot expect
the early onset of the pure perturbative regime
in the discussed diffractive processes: the
small interquark distances,
$\rho_{q\bar q} <0.2$ fm, start to dominate not earlier than  at
$Q^2 \simeq 100$ GeV$^2$, $W^2/Q^2 \simeq 10^7 $ in
$\gamma^*(Q^2)p\to\rho^0 p$ and
$Q^2 \simeq 50$ GeV$^2$, $W^2/Q^2 \simeq 10^6 $ in
$\gamma^*(Q^2)p\to\gamma^*(Q^2) p$.

\end{abstract}
%\pacs{12.39.-x, 11.55.Fv, 12.38.Bx, 14.40.Aq}

\section{Introduction}

In the present paper we continue a study of the onset of the perturbative
QCD regime in the processes initiated by virtual photon $\gamma^* (Q^2)$
along the lines of \cite{AMN,AMN_a}.
%A direct test of the QCD predictions for hard processes
%is an important problem.
One may address two types of hard processes which allow one to
study the transition regime
from the Strong-QCD physics to the
perturbative QCD physics of hard processes at accessible energies:
these are (i)
hard hadronic form factors \cite{ER,BL}, and  (ii)
pomeron-dominated reactions \cite{BFKL,Lip}.

An important problem which arises in the quantitative description of
hard processes is a numerical determination of the scale at which the
perturbative regime starts to dominate the amplitudes of interest.
Much attention to this subject has been paid in connection with the
elastic form factors \cite{ZhO,LSI}. In particular, the authors of \cite{ZhO}
advocated the viewpoint of an early applicability of pQCD at considerably low
$Q^2$, whereas in \cite{LSI}
the arguments in favour of the late onset of the
perturbative regime, at very large $Q^2\simeq 50$ GeV$^2$, were
given.  A quantitative study of \cite{AMN} confirmed the
arguments of ref. \cite{LSI} and showed that even at $Q^2$ as large as $20$
Gev$^2$ the form factor is still dominated by the soft non-perturbative
region of the $q\bar q$ kinematical configurations. In \cite{AMN,AMN_a}
the elastic pion and transition form factors were studied at
$0\leq Q^2\leq 100$ GeV$^2$ within an approach where both the truly
non-perturbative and perturbative contributions have been taken into
account. Namely, the form factor has been represented as a sum of the
triangle diagram (direct convolution of soft wave functions) and
diagrams with the one-gluon exchanges
(the convolution of hard one-gluon exchange kernel with soft wave functions).
The analysis of  form factors in the region of small $Q^2\sim 0 - 1$
GeV$^2$ allowed us to reconstruct  phenomenological soft wave functions of
pseudoscalar mesons and  soft photon. It was found that the
diagrams with the one-gluon exchange (the terms of the order of $\alpha_s$)
started to dominate the form factor only at $Q^2>50$ GeV$^2$, whereas at
smaller $Q^2$ the contribution of  non-perturbative triangle diagram was
substantial.

In this work we address diffractive
production by the hard photon, namely the reactions
$\gamma^*(Q^2)p\to\rho^0 p$ and $\gamma^*(Q^2)p\to\gamma^*(Q^2) p$.

One would have in mind the following physical picture of such processes
at asymptotically large $Q^2$: the virtual photon
produces $q\bar q$ pair at small distances (see Fig. 1), then
the quarks convert into vector meson (or outgoing photon)
with the $t$-channel emission of gluons:
the gluons at the top of the ladder are at relatively
small distances and thus a hard component of the pomeron (i.e. the BFKL
pomeron \cite{BFKL}) is selected.
The lower part of the gluon ladder is attached to the nucleon at small
momentum transfer, and thus the region of small transverse momenta is
selected at the bottom of the ladder.
Hence the most realistic is the following scenario of the whole reaction:
typical virtualities in the gluon ladder change from hard values selected
by the virtual photon in the hard quark-loop at the top of the ladder
to the soft values at the
bottom, and thus the whole $t$-channel behavior is determined by the
complicated object which is a superposition of the hard and the soft pomerons.
The details of the partition of the whole $t$-channel amplitude into the hard
and soft components depend on the values of $Q^2$ and specific details of the
selection of the hard component of the Pomeron by the quark loop. That is why a detailed
consideration of the hard upper quark-antiquark subprocess
based on realistic wave functions of the initial photon and the outgoing
vector meson is crucial for the understanding of the mechanism of the reaction
at large but finite $Q^2$.

Notice that the interest in better understanding of the diffractive
production mechanism is also motivated by the experimental resutls.
Experimentally a rather strong change of the $W$-dependence of the
cross section of the reactions initiated by the photon in different
regions of the photon virtuality has been observed:
Namely, the $W$ dependence of the cross section of the
photoproduction ($Q^2=0$) in the region $W\sim 10-200$ GeV$^2$ is rather flat,
similar to that in hadronic processes like $\pi p\to \pi p$.
However, the reaction $\gamma^*(Q^2)p \to Vp $  at $Q^2 \sim  10-20$ GeV$^2$
demonstrates an increase of the cross section with $W$ as $W^{2\Delta}$
where $\Delta \approx 0.3$.
An attractive possibility one might think of is to refer this growth to the
change of the pomeron regime from the Strong-QCD one at small $Q^2$ to
the BFKL one at large $Q^2$. However, for better understanding of this
phenomenon one needs a more detailed quantitative analysis of the
different ingredients of a complicated amplitude
of the diffractive production.

In this paper we concentrate on the quark block with a
coupled  photon, $\gamma^*(Q^2)$, which is responsible for a selection
of the region of small separations and thus of the hard pomeron component.
The quark block is determined by the convolution of photon wave functions
(the reaction $\gamma^*(Q^2)p\to\gamma^*(Q^2) p$) or
convolution of the photon and
$\rho$-meson wave functions (the reaction $\gamma^*(Q^2)p\to\rho^0 p$).

The light-cone wave function of photon
was found in ref. \cite{AMN_a} on the basis of experimental data
for the transition form factor $\gamma \gamma^*(Q^2) \to \pi^0$ at
$0 \leq Q^2 \leq 25$ GeV$^2$ \cite{g-pi}; this wave function
was successfully applied for the description of data on
$\gamma \gamma^*(Q^2) \to \eta$ and $\gamma \gamma^*(Q^2) \to \eta'$
\cite{g-pi,g-eta}. The analysis of the transition form factors has
been performed in terms of the double spectral representation over $q\bar q$
invariant mass squared, $M^2_{q\bar q} = \left ({m^2+k^2_{\perp} }
\right )/ \left (x(1-x) \right ) $
where $k^2_{\perp}$ and $x$
are the components of the light-cone quark momenta, and
$m$ is the constituent quark mass.
The photon wave function is represented as a
sum of two components which correspond to a direct production of the
$q\bar q$ pair with a point-like vertex $\gamma^* \to q\bar q$ at large
$M_{q\bar q}$, and to the production in the low-$M_{q\bar q} $ region
where the vertex has a nontrivial structure due to the soft $q\bar q$
interaction. The soft $q\bar q$ interaction yields an enhancement of the
contribution of low-masses and respectively large distances in the quark loop.

With the light-cone wave function of the photon at hand we can
calculate longitudinal and transverse polarization cross sections
$\gamma_L^*(Q^2)p\to\gamma_L^*(Q^2) p$ and
$\gamma_T^*(Q^2)p\to\gamma_T^*(Q^2) p$.

The onset of pQCD regime depends on the selection of
small interquark separations in a quark loop with $Q^2$.
So, it is important to test which fraction of the cross-section
as a function of $Q^2$ is actually gained at small transverse separations.
Therefore, in parallel with calculating the
$\gamma_{L,T}^*(Q^2)p\to\gamma_{L,T}^*(Q^2) p$
cross section which includes contribution of all transverse separations
in the quark-loop coupled to the BFKL pomeron,
we also calculate the part of this cross section which is actually gained at
small $q\bar q$ distances $\rho_{q\bar q}<0.2$ fm. Strictly speaking,
only this part of the cross section probes the BFKL pomeron, whereas
the cross section where the quarks in the loop are at large distances should be
rather coupled to the soft strong-QCD pomeron.

The results of our analysis show that in the region $Q^2 \sim 5 - 50$ GeV$^2$
the $\gamma^*(Q^2)p\to\gamma^*(Q^2) p$ cross section is dominated by the
domain of large transverse quark separations. The small quark
separations in the loop start to dominate the cross section only at
$Q^2 \geq 50$ GeV$^2$ and $W^2/Q^2\geq 10^6$.

This means that $\gamma^*(Q^2)$ selects small distances in the
$\gamma^*\gamma^*$-pomeron vertex approximately at the same values of
$Q^2$ as it was in the pion elastic and transition $\gamma \to
\pi, \; \eta,\; \eta'$ form factors.

The $\rho$-meson wave function necessary for the description of the reaction
$\gamma^*(Q^2)p\to\rho^0 p$ is not known promptly, so we have to use some
assumptions concerning the details of its behavior.
The low-$M_{q\bar q} $ component of the $\rho$-meson wave function is supposed
to be approximately the same as for a pion. The latter was
found in \cite{AMN} from fitting the data for pion form factor at
$0\leq Q^2 \leq 10$ GeV$^2$ \cite{pi}.
Moreover, the
$\rho$-meson low-$M_{q\bar q}$ wave function should be close to
photon's one due to the vector-meson dominance.

The high-mass component of the $\rho $-meson wave function needs a special
discussion: for its calculation, we cannot apply
exactly the same procedure as for the pion form factor.
In the analysis of pion form factor
\cite{AMN}, all the corrections of the order of $\alpha_s$ were taken
into account. To this end, the wave function of the pion was splitted into
soft and hard components, $\Psi^S$ and $\Psi^H$, in the following way
\begin{eqnarray}
\Psi^S \; {\rm is\; dominant\; at\;} M_{q\bar q}  < M_0, \qquad
\Psi^H\; {\rm is\; dominant\; at\;} M_{q\bar q} > M_0.
\end{eqnarray}
The parameter $M_0$, which separates the soft
and hard regions, is expected to be of the order of several GeV;
in \cite{AMN} the value $M_0=3$ GeV was chosen: this corresponds to
$|\vec{k}_0|^{-1}=\left (M_0^2/4-m^2 \right )^{-\frac{1}{2}}
 \simeq 0.15 \; {\rm fm}$ at $m\simeq 350$ MeV.
The separation of soft and hard wave functions was done in a simplest
way using the step-function:
\begin{eqnarray}
\Psi_{\pi}=  \Psi^S \theta(M_0-M_{q\bar q})+
\Psi^H \theta(M_{q\bar q}-M_0).
\end{eqnarray}
The hard component $\Psi^H$ is a convolution of the soft wave function
$\Psi^S$ and hard gluon exchange kernel $V^{\alpha_s}$:
\begin{eqnarray}
\Psi^H = V^{\alpha_s} \otimes  \Psi^S,
\end{eqnarray}
The splitting of the wave function into the soft and hard components
yields the following expansion  of the form factor over $\alpha_s$
\begin{eqnarray}
F=F^{SS}+2F^{SH}+O(\alpha_s^2)
\end{eqnarray}
where $F^{SS}$ is the soft form factor, and $F^{SH}$ is the soft-hard
contribution of the order $O(\alpha_s)$ which is determined by the
diagram with the hard gluon exchange, Eq. (3). In this way, including
the Sudakov form factor and renormalizing the quark mass,
we have taken into account all the terms of the order $O(\alpha_s)$
in Eq. (4).

A proper consideration of all the $\alpha_s$ terms in the reaction
$\gamma^*(Q^2)p\to\rho^ p$ is a much more
complicated task since such terms
appear not only as corrections to the $\rho$-meson vertex but also
as the next-to-leading order corrections to the BFKL-pomeron \cite{Fadin}
and pomeron-$q\bar q$ vertex. Such kind of analysis lies beyond the scope
of our present work. At the same time, the calculation of the
$\gamma^*(Q^2)p\to\rho^0 p$ amplitude with taking into account
only a part of the $\alpha_s$ corrections is not consistent.

In this situation, it seems reasonable to consider two extreme cases
of the behavior of the $\rho \to q\bar q$ vertex at large
$M^2_{q\bar q}$: first, $G_{\rho}(M^2_{q\bar q})\simeq M^{-2}_{q\bar q}$
and second $G_{\rho}(M^2_{q\bar q})= const\sim O(\alpha_s)$.
In both considered cases the dominance of the
region of small interquark separations
$\rho_{q\bar q} <0.2$ fm was not observed at least at
$Q^2 \le 50$ GeV$^2$ and $W^2/Q^2\geq 10^6$.

The paper is organized as follows:

In Section II we discuss the kinematics of the reaction,
the Lorentz structure of the amplitudes
and some principal points of the spectral representations. Section III presents
a detailed calculation of the amplitudes in the framework of
the spectral representations over the invariant $q\bar q$-mass. The
numerical results are discussed in Section IV. In Conclusion we
summarize the main results and present our viewpoint on the Strong-QCD pomeron
and the electoproduction of vector mesons at $Q^2 < 100$
GeV$^2$.

\section{Kinematics of the reactions and the Lorentz
structure of the amplitudes}

In this Section we describe the kinematics of the reaction and outline the
principle points in the dispersive representation of the amplitude.
Further technical details and final analytical expressions are
given in the next Section.

We use the following notation for the
four-vector of the particles involved:
  the initial virtual photon four-momentum $q$, the final virtual
photon (or vector meson) four-momentum $q'$,
the target nucleon momentum $p_N$,
and the outgoing nucleon momentum $p'_N$. The following momentum
conservations are imposed:
\begin{equation}
q'=q-\kappa, \quad p'_N=p_N+\kappa,\quad  q+p_N=q'+p'_N.
\end{equation}
The reaction is characterized by three  independent kinematical variables
$Q^2=-q^2$, $W=(p_Nq)/m_N$, and $\kappa^2$, since
$p_N^2=p'^2_N=m_N^2$ and  $q'^2=\mu_V^2$ for
$\gamma^*(Q^2)p\to V p$ or $q'^2=-Q^2$ for
$\gamma^*(Q^2)p\to\gamma^*(Q^2) p$ .

We are interested in the kinematics when
\begin{equation}
\label{kin}
Q^2 {\rm \;\; is\;\; large},  \; \;
 W^2/Q^2\gg 1,\qquad {\rm and\;} \; \kappa^2\to 0.
\end{equation}
It is convenient to describe the process in the reference frame where
the target nucleon is at rest $p_N=(m_N, 0,0)$
and photon  fastly moving along the longitudinal axis $z$.
The kinematical conditions (\ref{kin}) then imply
that $q_z\to\infty$ and omitting the $1/q_z^2$ terms
we come to the following expressions for the
momentum and polarization vectors of the incoming photon
\begin{eqnarray}
q=(q_z-\frac{Q^2}{2q_z}, 0,q_z),
\qquad
\epsilon _{L}^{(\gamma)}(Q^2)=\frac{1}{iQ} (q_z,0,q_z-\frac{Q^2}{2q_z}),
\qquad
\epsilon _{T}^{(\gamma)}(Q^2)=(0,\vec e^{\;(\gamma)}_\perp,0).
\end{eqnarray}
For the outgoing vector meson in the reaction $\gamma^*(Q^2)p\to V p$,
one has
\begin{eqnarray}
q'=(q'_z+\frac{\mu^2_{\perp}}{2q'_z},-\vec\kappa_{\perp},q'_z),
\qquad
\epsilon _{L}^{(V)}(\mu^2)=\frac{1}{\mu_{\perp}}
(q'_z,0,q'_z+\frac{\mu_{\perp}^2}{2q'_z}),
\qquad
\epsilon _{T}^{(V)}=(0,\vec e^{\; (V)}_\perp,0).
\end{eqnarray}
where $\mu^2_\perp=\mu^2+\kappa_\perp^2$ and
$\vec e^{\; (V)}_\perp\vec\kappa_\perp=0$.
In the limit
$\kappa^2_{\perp}/m_N \to 0$, the
momenta $q'_z$ and $\kappa$ are equal to
\begin{equation}
q'_z=q_z -\frac{\mu_{\perp} ^2+Q^2}{2q_z}, \qquad
\kappa =q-q' =(0,\vec \kappa_{\perp},
\frac{\mu_{\perp} ^2+Q^2}{2q_z}).
\end{equation}
Similar expressions determine the momentum and polarization vectors
of the outgoing photon in the reaction
$\gamma^*(Q^2)p\to\gamma^*(Q^2) p$: one needs to substitute
$\mu_V^2 \to -Q^2 $. Because of that we present below the formulas
for the vector meson production only.

Instead of the nucleon momentum $p_N$,
 it is convenient to characterize the
reaction under the kinematical conditions (\ref{kin}) by the
vector
\begin{equation}
\label{n}
n=\frac{1}{q_0+q_z}(1,0,-1).
\end{equation}
 The vectors $q$ and $n$ determine the plane for
longitudinal polarization, whereas the transverse plane is determined by
transverse components of the vector $\kappa$.
Notice that $n^2=0$ and $qn=-1$.

The amplitude of the reaction can be written in the form
\begin{eqnarray}
A=\epsilon^{(\gamma)}_\mu A_{\mu\nu}(q,q',n)\epsilon^{(V)}_\nu,
\end{eqnarray}
where, due to the gauge invariance, the tensor $A_{\mu\nu}$
satisfies the relations
$q^\mu A_{\mu\nu}(q,q',n)=q'^\nu A_{\mu\nu}(q,q',n)=0$,
thus having the following Lorentz structure:
\begin{eqnarray}
\label{ampl}
\nonumber
A_{\mu\nu}(q,q',n)&=&-(g_{\mu\nu}-\frac{n_\mu q_\nu}{nq}
-\frac{q'_\mu n_\nu}{nq'}
+\frac{qq'}{(nq)\;(nq')}n_\mu n_\nu)\;A_T(q^2,q'^2,nq,\kappa^2)\\
&&+(q^2 n_\mu-nq\cdot q_\mu)(q'^2 n_\nu-nq'\cdot q'_\nu)
\;A_L(q^2,q'^2,nq,\kappa^2)+O(\kappa_\mu, \kappa_\nu),
\end{eqnarray}
and $O(\kappa_\mu, \kappa_\nu)$ stands for other possible tensor
structures transversal with respect to $q_\mu$ and $q'_\nu$ and
proportional to $\kappa$.
In the limit  $\kappa\to 0$ they do not contribute
to the cross section and are omitted.

The cross sections are connected with the introduced amplitudes as
follows:
\begin{eqnarray}
\frac{d\sigma(\gamma^*_T(Q^2)p\to V_T p)}{d(-\kappa^2)}
=|A_T|^2; \qquad
\frac{d\sigma(\gamma^*_{L}(Q^2)p\to V_{L} p)}{d(-\kappa^2)}
=Q^2\mu^2_V|A_L|^2,
\end{eqnarray}
where we have omitted the terms proportional to $\kappa^2$.
The invariant amplitudes $A_{L,T}$ are connected with
$A_{\mu\nu}$ through the following relations:
\begin{eqnarray}
A_T=-\epsilon^{(\gamma)}_{T\mu}(Q^2)A_{\mu\nu}(q,q',n)
\epsilon^{(V)}_{T\nu}(\mu^2_V), \qquad
A_L=\frac{1}{i\mu_VQ}
\epsilon^{(\gamma)}_{L\mu}(Q^2)A_{\mu\nu}(q,q',n)
\epsilon^{(V)}_{L\nu}(\mu^2_V).
\end{eqnarray}

One can write the spectral representations for the
invariant amplitudes $A_{L,T}$ as follows:
\begin{eqnarray}
A_{L,T}(q^2,q'^2)=\int \frac{dM^2_{q\bar q} \;
 dM'^2_{q\bar q}}{\pi^2}
\frac{{\rm disc}_{M^2_{q\bar q}}  \;{\rm disc}_{ M'^2_{q\bar q}} \;
A_{L,T}(M^2_{q\bar q}, M'^2_{q\bar q})}{(M^2_{q\bar q}-q^2)
( M'^2_{q\bar q}-q'^2)}.
 \end{eqnarray}
 In order to obtain
the double spectral densities ${\rm disc}_{M^2_{q\bar q} }
\;{\rm disc}_{ M'^2_{q\bar q}}\;
A_{L,T}(M^2_{q\bar q},M'^2_{q\bar q})$ we must consider the process
with the off-shell incoming
and outgoing particles.  Namely, we have the following off-shell
momenta
\begin{equation} q'\to P', \qquad q\to P,
\end{equation}
where
\begin{equation}
P^2=M^2_{q\bar q}, \qquad P\;'^2=
M'^2_{q\bar q}, \qquad  ( P\;'-P)^2=\kappa ^2,
\end{equation}
but
\begin{equation}
P-P\; ' \ne \kappa,
\end{equation}
if we take into account the terms of the order of $1/q_z$
and omit $O(1/q^2_z)$-terms.
The components of the off-shell incoming $P$ and outgoing $P'$
momenta can be chosen as follows:
\begin{equation}
P=(q_z+\frac{M^2_{q\bar q}}{2q_z}, 0,q_z), \qquad
P\; '=(q_z+
\frac{M'^2_{q\bar q}+\kappa^2_{\perp}}{2q_z},-\vec\kappa_{\perp} ,q_z).
\end{equation}

The off-shell amplitude $A_{\mu\nu}(P,P',n)$ has a
 similar decomposition in
terms of $P,P',n$ as has the amplitude $A_{\mu\nu}(q,q',n)$
of Eq. (\ref{ampl}) in terms of
$q,q',n$.

Let us introduce the polarization vectors of the off-shell
vector particles. The corresponding
longitudinal polarization vectors are
\begin{equation}
\epsilon _{L}^{(P)} (M^2_{q\bar q})=\frac{1}{\sqrt { M^2_{q\bar q} }}
(q_z,0,q_z+\frac{M^2_{q\bar q}}{2q_z}),
\qquad
\epsilon _{L}^{(P\; ')}
(M'^2_{q\bar q})=\frac{1}{\sqrt {M'^2_{q\bar q}+\kappa^2_{\perp}}}
(q_z,0,q_z+\frac {M'^2_{q\bar q}+\kappa^2_{\perp}}{2q_z}).
\end{equation}
The transverse polarization vectors of the off-shell particles
are equal to the transverse polarization vectors of the incoming
and outgoing particles.

The amplitudes $A_{L,T}$ are connected with $A_{\mu\nu}$ as follows:
\begin{eqnarray}
A_T=-\epsilon^{(P)}_{T\mu}A_{\mu\nu}(P,P',n)\epsilon^{(P')}_{T\nu}, \qquad
A_L=\frac{1}{\sqrt{M^2_{q\bar q}
M'^2_{q\bar q}}}\epsilon^{(P)}_{L\mu}A_{\mu\nu}(P,P',n)
\epsilon^{(P')}_{L\nu},
\end{eqnarray}
where the factor $\sqrt{M^2_{q\bar q}M'^2_{q\bar q}}$
is just the following expression:
\begin{equation}
\epsilon^{(P)}_{L\mu}
(P^2 n_\mu-nP\cdot P_\mu)(P'^2 n_\nu-nP'\cdot P'_\nu)
\epsilon^{(P')}_{L\nu}=\frac{\sqrt{M^2_{q\bar q}}M'^2_{q\bar q}}
{\sqrt{M'^2_{q\bar q}+\kappa_\perp^2}} \simeq\sqrt{M^2_{q\bar q}
M'^2_{q\bar q}}.
\end{equation}
Hence,
\begin{eqnarray}
\nonumber
{\rm disc}_{M^2_{q\bar q}} {\rm disc}_
{M'^2_{q\bar q}} A_T&=&{\rm disc}_{M^2_{q\bar q}}
 {\rm disc}_{M'^2_{q\bar q}}
\epsilon^{(P)}_{T\mu}A_{\mu\nu}(P,P',n)\epsilon^{(P')}_{T\nu},
\\
{\rm disc}_{M^2_{q\bar q}} {\rm disc}_{M'^2_{q\bar q}}A_L&=&{\rm
disc}_{M^2_{q\bar q}}  {\rm disc}_{M'^2_{q\bar q}}
\epsilon^{(P)}_{L\mu}A_{\mu\nu}(P,P',n)\epsilon^{(P')}_{L\nu}
\frac{1}{\sqrt{M^2_{q\bar q}M'^2_{q\bar q}}}.
\end{eqnarray}
Finally, recall that the double spectral density of the amplitude
$A_{\mu\nu}(P,P',n)$ is calculated by placing all particles in the
intermediate mass-on-shell states.

Notice that the amplitude $A_T$ can be
directly isolated from $A_{\mu\nu}$.
Namely, setting $\mu,\nu=a,b=1,2$ and isolating the term
$-g_{\mu\nu} \to \delta_{ab}$, one obtains $A_T$
\begin{equation}
A_{ab}(P,P',n)=\delta_{ab}A_T+O(\kappa_a\kappa_b),
\end{equation}
since $\vec \kappa$ is the only vector with the components in the
transverse plane. In the next Section we perform a detailed
consideration of $A_{L,T}$.

\section{Spectral representation of the amplitudes}

The amplitudes of the reactions are given by the diagrams of Fig 1.
First, we shall consider separately the upper quark-loop block
and then take into account its interaction with a nucleon through
the Pomeron exchange. We shall discuss
in parallel the cases of the transverse and longitudinal
polarizations of the
initial photon and outgoing vector meson (photon). As before,
 we concentrate ourselves on the reaction
$\gamma^*(Q^2)p\to V p$ keeping in mind that for
$\gamma^*(Q^2)p\to\gamma^*(Q^2) p$ the formulas are written
analogously.

\subsection{The block of the photon and vector meson  interaction
with the BFKL Pomeron }

The Pomeron is attached to the photon and the vector meson through the quark
loop diagrams of Fig. 1. We consider separately the cases
when both Reggeized gluons
are attached to the same and to different quarks in the loop.

\subsubsection{The gluon ladder attached to a single constituent}

The diagram for this subprocess is shown in Fig 2a. The analytical spectral
representation for this quark-loop diagram has the following structure
\begin{equation}
A^{I}_{L,T}=\int\frac{dM_{q\bar q}^2}\pi G_\gamma(M_{q\bar q}^2)
\frac{d\Phi_2(P;k_1,k_2)}{M_{q\bar q}^2-q^2-i0}
\frac{dM''^2_{q\bar q}}\pi\frac{d\Phi_1(P'';k_1'',k_2)}
{M''^2_{q\bar q}-(q-k)^2-i0}
\frac{dM'^2_{q\bar q}}\pi\frac{d\Phi_1(P';k_1',k_2)}
{M'^2_{q\bar q}-(q-\kappa)^2-i0}
G_V(M'^2_{q\bar q})g^2(-1)S^I_{L,T},
\label{start1}
\end{equation}
where $P^2=M_{q\bar q}^2$, $P'^2=M'^2_{q\bar q}$, $P''^2=M''^2_{q\bar
q}$ are the invariant masses squared of the $q\bar q$ pairs in the
intermediate states and the corresponding phase space factors read
\begin{eqnarray}
d\Phi_2(P;k_1,k_2)&=&\frac 12 \frac{d^3k_1}{(2\pi)^3 2 k_{10}}
\frac{d^3k_2}{(2\pi)^3 2 k_{20}} (2\pi)^4\delta^4(P-k_1-k_2)
\nonumber \\
&=&\frac 1{(4\pi)^2}\frac{dx_1dx_2}{x_1x_2}\delta(1-x_1-x_2)
d^2k_{1\perp} d^2k_{2\perp}\delta(\vec{k}_{1\perp}+\vec{k}_{2\perp})
\delta\left(M_{q\bar q}^2-\frac{m^2_{1\perp}}{x_1}
-\frac{m^2_{2\perp}}{x_2}\right),
\nonumber \\
d\Phi_1(P'';k_1'',k_2)&=&\frac 12 \frac{d^3k_1''}{(2\pi)^3 2
k_{10}''} (2\pi)^4\delta^4(P''-k_1''-k_2) \nonumber \\
&=&\pi\frac{dx_1''}{x_1''}\delta(1-x_1''-x_2)d^2k_{1\perp}''
\delta(\vec{k}_{1\perp}''+\vec{k}_\perp+\vec{k}_{2\perp})
\delta\left(M''^2_{q\bar q}+k^2_\perp
-\frac{m''^2_{1\perp}}{x_1''}-\frac{m^2_{2\perp}}{x_2}\right),
\nonumber \\
d\Phi_1(P';k_1',k_2)&=&\frac 12 \frac{d^3k_1'}{(2\pi)^3 2 k_{10}'}
(2\pi)^4\delta^4(P'-k_1'-k_2)
\nonumber \\
&=&\pi\frac{dx_1'}{x_1'}\delta(1-x_1'-x_2)d^2k_{1\perp}'
\delta(\vec{k}_{1\perp}'+\vec{\kappa}_\perp+\vec{k}_{2\perp})
\delta\left(M'^2_{q\bar q}+\kappa^2_\perp
-\frac{m'^2_{1\perp}}{x_1'}-\frac{m^2_{2\perp}}{x_2}\right).
\label{phas-vol}
\end{eqnarray}
Here we have taken into account that $k_z$ is small:
the integration over $k_z$ is performed  enclosing the
integration contour over the pole
$(M''^2_{q\bar q}+Q^2-2q_zk_z-i0)^{-1}$, that is equivalent to
$(M''^2_{q\bar q}+Q^2-2q_zk_z-i0)^{-1}
\to \; -i\pi \delta (M''^2_{q\bar q}+Q^2-2q_zk_z)$.
As a result, we find
\begin{eqnarray}
A^{I}_{L,T}&=&\frac 1{4\pi} \int^{1}_{0}\frac{dx}{x(1-x)^3}
\int \frac{d^2k_{2\perp}}{(2\pi)^2}
dk_z \frac{G_\gamma(M_{q\bar q}^2)}{M_{q\bar q}^2+Q^2}\;
\frac 1{M''^2_{q\bar q}+Q^2-2q_zk_z-i0}\;
\frac{G_V(M'^2_{q\bar q})}
{M'^2_{q\bar q}-\mu^2_V} g^2(-1)S^I_{L,T}
\nonumber \\
&=&\int^{1}_{0} \frac{dx}{x(1-x)^3}
\int\frac{d^2k_{2\perp}}{(4\pi)^2}
\frac{G_\gamma(M_{q\bar q}^2)}{M_{q\bar q}^2+Q^2}\;
\frac{G_V(M'^2_{q\bar q})}{M'^2_{q\bar q}-\mu^2_V}\;
\frac {-ig^2}{2q_z} S^I_{L,T},
\label{A^I}
\end{eqnarray}
where
\begin{eqnarray}
M_{q\bar q}^2&=&\frac{m^2+k^2_{2\perp}}{x(1-x)},\quad
M''^2_{q\bar q}=
\frac{m^2+(\vec{k}_{2\perp}+x\vec{k}_\perp)^2}{x(1-x)},\quad
M'^2_{q\bar q}=
\frac{m^2+(\vec{k}_{2\perp}+x\vec{\kappa}_\perp)^2}{x(1-x)},\quad
x\equiv x_2.
\label{smas1}
\end{eqnarray}

The spin factor $S^I_{L,T}$ appears due to
a decomposition of the trace $S^{I}_{\mu\nu}$
corresponding to the quark loop of Fig. 2a
\begin{eqnarray}
\label{sp1}
S^{I}_{\mu\nu}=Tr \left( \gamma_\nu(\hat{k}_1'+m) \hat{n}
(\hat{k}_1''+m) \hat{n}(\hat{k}_1+m)\gamma_\mu
(-\hat{k}_2+m)\right),
\end{eqnarray}
where $\hat{n}$, being determined by (\ref{n}),
stands for the vertex of the reggeized gluon-quark
coupling \cite{FGL}. Let us
stress once more that all quarks in this expression
are mass-on-shell, whereas the photon and the vector meson momenta are
mass-off-shell, such that $P^2=M^2_{q\bar q}$ and $P'^2=M'^2_{q\bar q}$.  As
we have discussed in the previous Section, to obtain the transverse
amplitude, i.e. $S^I_T$, we must set $\mu,\nu=a,b=1,2$ and isolate the
structure proportional to $\delta_{ab}$.  This procedure yields $S^I_T$ in
the form
\begin{equation} S^I_T= -8 \frac{(1-x)}x \left [m^2+\left (
1-2x(1-x) \right ) \vec k_{2\perp}\vec k'_{2\perp} \right ],
\end{equation}
where $\vec k'_{2\perp}=\vec k_{2\perp}+x\vec\kappa_\perp$. Note that the
last term in the square brackets is not small, playing significant role
in the determination of $A_T$.

The spin-factor for longitudinal polarization can be
calulated performing the convolution with longitudinal
polarization vectors of the off-shell photon and vector meson as follows:
\begin{equation}
S_L^I= \frac {
\epsilon^{(V)}_{L\ \beta}(P'^2)
Tr \left( \gamma_\beta(\hat{k}_1'+m) \hat{n}
(\hat{k}_1''+m) \hat{n}(\hat{k}_1+m)\gamma_\alpha
(-\hat{k}_2+m)\right)\epsilon^{(\gamma)}_{L\ \alpha}(P^2)}
{P^2P'^2
\left(\epsilon^{(V)}_{L}(P'^2)n\right)
\left(\epsilon^{(\gamma)}_{L}(P^2)n\right)}\ .
\label{sp1l}
\end{equation}
The calculation of this trace yields the following result
\begin{eqnarray}
S_L^I&=&
-32\; \frac{(1-x)}x\;
\frac{x^2(1-x)^2}{m^2+{k'}_{2\perp}^2+x(1-x)\kappa_\perp^2}
\left[{m^2+k'^2_{2\perp}-(x-1/2){\vec k'}_{2\perp}\vec\kappa_\perp}
\right] \to -32x(1-x)^3.
\label{ssspI}
\end{eqnarray}
Here we take into account that the last term in the square brackets
gives a negligible contribution into the BFKL amplitude.

\subsubsection{The gluon ladder attached to both constituents}

This subprocess is displayed in Fig 2b. Likewise, the triple spectral
representation for the corresponding amplitude takes a form:
\begin{equation}
A^{II}_{L,T}=\int\frac{dM_{q\bar q}^2}\pi G_\gamma(M_{q\bar q}^2)
\frac{d\Phi_2(P;k_1,k_2)}{M_{q\bar q}^2-q^2-i0}
\frac{dM''^2_{q\bar q}}\pi\frac{d\Phi_1(P'';k_1',k_2)}
{M''^2_{q\bar q}-(q-k)^2-i0}
\frac{dM'^2_{q\bar q}}\pi\frac{d\Phi_1(P';k_1',k_2')}
{M'^2_{q\bar q}-(q-\kappa)^2-i0}
G_V(M'^2_{q\bar q})g^2(-1)S^{II}_{L,T}.
\end{equation}
Now the phase space factors read
\begin{eqnarray}
d\Phi_2(P;k_1,k_2)&=&
\frac 1{(4\pi)^2}\frac{dx_1dx_2}{x_1x_2}\delta(1-x_1-x_2)
d^2k_{1\perp} d^2k_{2\perp}\delta(\vec{k}_{1\perp}+\vec{k}_{2\perp})
\delta\left(M_{q\bar q}^2-\frac{m^2_{1\perp}}{x_1}
-\frac{m^2_{2\perp}}{x_2}\right),
\nonumber \\
d\Phi_1(P'';k_1',k_2)&=&
\pi\frac{dx_1'}{x_1'}\delta(1-x_1'-x_2)d^2k_{1\perp}'
\delta(\vec{k}_{1\perp}'+\vec{k}_\perp+\vec{k}_{2\perp})
\delta\left(M''^2_{q\bar q}+k^2_\perp
-\frac{m'^2_{1\perp}}{x_1'}-\frac{m^2_{2\perp}}{x_2}\right),
\nonumber \\
d\Phi_1(P';k_1',k_2')&=&
\pi\frac{dx_2'}{x_2'}\delta(1-x_1'-x_2')d^2k_{2\perp}'
\delta(\vec{k}_{1\perp}'+\vec{\kappa}_\perp+\vec{k}_{2\perp}')
\delta\left(M'^2_{q\bar q}+\kappa^2_\perp
-\frac{m'^2_{1\perp}}{x_1'}-\frac{m'^2_{2\perp}}{x_2'}\right).
\end{eqnarray}
This time the expressions for the invariants
$M_{q\bar q}^2,M''^2_{q\bar q},M'^2_{q\bar q}$ take a form:
\begin{equation}
M_{q\bar q}^2=
\frac{m^2+k^2_{2\perp}}{x(1-x)},\quad
M''^2_{q\bar q}=
\frac{m^2+(\vec{k}_{2\perp}+(1-x)\vec{k}_\perp)^2}{x(1-x)},\quad
M'^2_{q\bar q}=
\frac{m^2+(\vec{k}_{2\perp}+\vec{k}_\perp-x\vec{\kappa}_\perp)^2}{x(1-x)},
\quad x\equiv x_1.
\label{smas2}
\end{equation}
 Performing the integration over $k_z$ we come to the expression:
\begin{equation}
A^{II}_{q\bar q}=\int^{1}_{0} \frac{dx}{x^2(1-x)^2}
\int\frac{d^2k_{2\perp}}{(4\pi)^2}
\frac{G_\gamma(M_{q\bar q}^2)}{M_{q\bar q}^2+Q^2}\;
\frac{G_V(M'^2_{q\bar q})}{M'^2_{q\bar q}-\mu^2_V}\;
\frac{-ig^2}{2q_z}S^{II}_{L,T}.
\end{equation}
The quark loop trace for the diagram of Fig. 2b reads
\begin{equation}
\label{sp2}
S^{II}_{\mu\nu}= Sp \left( \gamma_\nu(\hat{k}_1'+m) \hat{n}
(\hat{k}_1+m)\gamma_\mu (-\hat{k}_2+m)\hat{n}
(-\hat{k}_2'+m)\right).
\end{equation}

Again, setting $\mu,\nu=a,b=1,2$ and isolating the factor
proportional to the
$\delta_{ab}$ yield a simple expression:
\begin{equation}
S_T^{II}=
8\left[ m^2+
\left( 1-2x(1-x) \right )
{\vec k_{2\perp}}{\vec k''_{2\perp}}\right ],
\label{ssp2}
\end{equation}
where we have introduced the vector
\begin{equation}
\label{k''}
\vec k''_\perp=\vec{k}_{2\perp}+\vec{k}_\perp-x\vec{\kappa}_\perp.
\end{equation}
For longitudinal polarization the spin factor reads
\begin{eqnarray}
\label{sp2l}
S_L^{II}&=&\frac {
\epsilon^{(V)}_{L\ \beta}(P'^2)
Sp \left( \gamma_\beta(\hat{k}_1'+m) \hat{n}
(\hat{k}_1+m)\gamma_\alpha (-\hat{k}_2+m)\hat{n}
(-\hat{k}_2'+m)\right)\epsilon^{(\gamma)}_{L\ \alpha}(P^2)}
{P^2P'^2
\left(\epsilon^{(V)}_{L}(P'^2)n\right)
\left(\epsilon^{(\gamma)}_{L}(P^2)n\right)}\ .
\end{eqnarray}
which, after calculating the trace, takes the following form:
\begin{equation}
S_L^{II}=32\; \frac{x^2(1-x)^2}
{m^2+{k''_2}^2}\;
\left[ m^2+k''^2_{2\perp}
+(x-1/2)\vec\kappa_\perp \vec k''_{2\perp}\right ] \to 32x^2(1-x)^2.
\label{sssp2}
\end{equation}
Here, as in (\ref{ssspI}), we take into account that the last term in
the square brackets gives a small contribution into the BFKL amplitude.

The vertices $G_\gamma(M_{q\bar q}^2)$ and $G_V(M'^2_{q\bar q})$ which
are used for the calculation are shown in Fig. 3.

\subsection{BFKL-pomeron couplings to the quark loop and nucleon}

Now we must attach the quark loop and target nucleon to
 the BFKL-pomeron amplitude; we face several possible
scenarios for this attaching.

\subsubsection{Pomeron-nucleon coupling}

We consider two possible scenarios for
 the Pomeron attaching to the
nucleon target:\\
(i) the BFKL Pomeron is transformed first to the
soft Pomeron and then this soft Pomeron is
attached to the nucleon,  \\
(ii) the BFKL is directly attached to the nucleon.

The momentum transfer to the Pomeron $\kappa$ is not large and there is
no selection of the small separations along the gluon
ladder: while moving from the quark block to the nucleon, the distances
between the $t$ channel gluons in the impact parameter space are increasing.

(i) {\bf Soft-Pomeron-nucleon coupling}.
If the ladder is rather long (i.e. at large $W$),
the distances between the
gluons become the normal hadronic distances, and in
this case the Pomeron is no
longer the perturbative BFKL  but is rather in a soft regime. This
soft Pomeron is then attached to the nucleon, and we use the standard
soft-Pomeron-nucleon coupling which we denote as
$\tilde g F_{PNN}(\kappa^2_\perp)$.
Within this scenario and following to the prescription of refs.
\cite{BFKL,Lip,ryskin,forshaw}, we write the amplitude
of Fig. 1a for the case,
when both Reggeized gluons are attached to the same
quark (or antiquark):
\begin{eqnarray}
A_{L,T}^{BFKL-q}(\gamma^* p\to Vp)&=&-\frac i2\int^{1}_{0}
\frac{dx}{x^2(1-x)^2}
\int\frac{d^2k_{2\perp}}{(2\pi)^2}
\frac{d^2k_\perp}{(2\pi)^2}\;
\Psi_\gamma(x,k^2_{2\perp})\;
\Psi_V\left(x, \left( \vec{k}_{2\perp}+x\vec{\kappa}_\perp \right)^2
\right)
\nonumber \\
&\times&\frac{x}{1-x} S^I_{L,T}g^2 \int^{\infty}_{-\infty}
\frac{d\nu\;\;\nu^2}{(\nu^2+\frac 14)^2}
\left(\frac{W^2}{Q^2+\mu^2_V}\right)^{\omega(\nu)}
\int d^2\rho_1 d^2\rho_2
e^{i\vec{k}_\perp(\vec{\rho}_1-\vec{\rho}_2)}
e^{i\vec{\kappa}_\perp\vec{\rho}_2}
\nonumber \\
&\times&\left(
\left[\frac{(\vec{\rho}_1-\vec{\rho}_2)^2}{\rho^2_1\rho^2_2}\right]
^{\frac 12 +i\nu}
-\left[\frac 1{\rho^2_1}\right]^{\frac 12 +i\nu}
-\left[\frac 1{\rho^2_2}\right]^{\frac 12 +i\nu}
\right)
\tilde g F_{PNN}(\kappa^2_\perp),
\label{tot1}
\end{eqnarray}
where
\begin{equation}
\Psi_\gamma(x,k^2_{2\perp})=
\frac{G_\gamma(M_{q\bar q}^2)}{M_{q\bar q}^2+Q^2},\quad
\Psi_V\left(x, \left( \vec{k}_{2\perp}+x\vec{\kappa}_\perp\right)^2\right)
=\frac{G_V(M'^2_{q\bar q})}{M'^2_{q\bar q}-\mu^2_V}\ ,
\end{equation}
and the invariant masses
$M^2_{q\bar q}$, $M'^2_{q\bar q}$ are given by
(\ref{smas1}).

The variables $\vec{\rho}_1$ and $\vec{\rho}_2$ are the gluon coordinates
in the impact parameter space.
The energy dependence of the amplitude is given by the function $\omega(\nu)$:
\begin{equation}
\omega(\nu)=\frac{2\alpha_s C_A}\pi
Re\left(
\frac{\Gamma'(1)}{\Gamma(1)}
-\frac{\Gamma'\left( \frac 12+i\nu\right)}
{\Gamma\left( \frac 12+i\nu\right)}
\right),
\end{equation}
with $C_A=N_c=3$ and $\Gamma(z)$ the Euler $\Gamma$-function.

Likewise, the amplitude of Fig. 1b, when the gluons are attached to
quark and antiquark, takes the form
\begin{eqnarray}
A_{L,T}^{BFKL-q\bar q}(\gamma^*p\to Vp)&=&-\frac i2\int^{1}_{0}
\frac{dx}{x^2(1-x)^2}
\int\frac{d^2k_{2\perp}}{(2\pi)^2}
\frac{d^2k_\perp}{(2\pi)^2}\;
\Psi_\gamma(x,k^2_{2\perp})\;
\Psi_V\left(x,\left( \vec{k}_{2\perp}+\vec{k}_\perp-x\vec{\kappa}_\perp
\right)^2\right)
\nonumber \\
&\times&g^2S^{II}_{L,T}\int^{\infty}_{-\infty}
\frac{d\nu\;\;\nu^2}{(\nu^2+\frac 14)^2}
\left(\frac{W^2}{Q^2+\mu^2_V}\right)^{\omega(\nu)}
\int d^2\rho_1 d^2\rho_2
e^{i\vec{k}(\vec{\rho}_1-\vec{\rho}_2)}
e^{i\vec{\kappa}_\perp\vec{\rho}_2}
\nonumber \\
&\times&\left(
\left[\frac{(\vec{\rho}_1-\vec{\rho}_2)^2}{\rho^2_1\rho^2_2}\right]
^{\frac 12 +i\nu}
-\left[\frac 1{\rho^2_1}\right]^{\frac 12 +i\nu}
-\left[\frac 1{\rho^2_2}\right]^{\frac 12 +i\nu}
\right)
\tilde g F_{PNN}(\kappa^2_\perp),
\label{tot2}
\end{eqnarray}
where
\begin{equation}
\Psi_\gamma(x,k^2_{2\perp})=
\frac{G_\gamma(M_{q\bar q}^2)}{M_{q\bar q}^2+Q^2},\quad
\Psi_V\left(x, \left( \vec{k}_{2\perp}+\vec k_\perp - x\vec{\kappa}_\perp\right)^2\right)
=\frac{G_V(M'^2_{q\bar q})}{M'^2_{q\bar q}-\mu^2_V}\ ,
\end{equation}
and the invariant masses $M^2_{q\bar q}$ and
$M'^2_{q\bar q}$ are given by (\ref{smas2}).

The nucleon-Pomeron coupling $\tilde g F_{PNN}(\vec{\kappa}^2_\perp)$
can be well approximated assuming
$F_{PNN}(\kappa^2_\perp)= e^{-B\kappa^2_\perp}$ with
$B\simeq 2.5\; GeV^{-2}$.

The total amplitude is obtained summing the amplitudes of all
subprocesses,  it takes the form
\begin{equation}
A^{BFKL}(\gamma^* p\to Vp)=
A_{L,T}^{BFKL-q}(\gamma^* p\to Vp)
+A_{L,T}^{ BFKL-\bar q}(\gamma^* p\to Vp)
+2\;A_{L,T}^{BFKL-q\bar q}(\gamma^* p\to Vp)\ .
\end{equation}

Some important cancellations occur in the total amplitude: namely,
the terms in the BFKL amplitude proportional
to $\rho_1$ or $\rho_2$ separately cancel each other.
In addition, the term proportional to $(\rho_1-\rho_2)^2$
in $A^{BFKL-q}_{L,T}$
vanishes, since the $k_\perp$ integration yields
$\delta(\rho_1-\rho_2)$.

Finally, we come to the following representation of the
total amplitude of the diffractive vector meson production
\begin{eqnarray}
A^{BFKL}_{L,T}(\gamma^*p\to V p)&=&-i
\int^{1}_{0} dx
\int^{\infty}_{0}
\frac{d\nu\;\;\nu^2}{(\nu^2+\frac 14)^2}
\left(\frac{W^2}{Q^2+\mu^2_V}\right)^{\omega(\nu)}
\nonumber \\
&\times&
\int d^2\rho_1 d^2\rho_2
\exp\left[{i\vec\kappa_\perp(\vec\rho_1x+\vec\rho_2(1-x))}\right]
\left[\frac{(\vec{\rho}_1-\vec{\rho}_2)^2}{\rho^2_1\rho^2_2}\right]
^{\frac 12 +i\nu}
\nonumber \\
&\times&
\int
\frac{d^2k''_{2\perp}}{(2\pi)^2}
\frac{d^2k_{2\perp}}{(2\pi)^2}
\Psi_\gamma(x,k^2_{2\perp})\Psi_V(x,{k''^2_{2\perp}})
\exp\left[{i(\vec k''_{2\perp}-\vec k_{2\perp})(\vec\rho_1-\vec\rho_2)}\right]
S_{L,T}^{II}
C e^{-B\kappa^2_\perp}\ .
\label{totfin1}
\end{eqnarray}
where $k''_{2\perp}$ is given by (\ref{k''}), and $C=g^2\tilde g$ plays
a role of the normalization constant.

It is convenient to transform the
final expression for the photon and vector meson  wave functions to
the coordinate representation as follows:
\begin{eqnarray}
A^{BFKL}_{L,T}(\gamma^* p\to Vp)&=&
-\frac{i}{(2\pi)^2}\; C e^{-B\kappa^2_\perp}
\int^{1}_{0} \frac{dx}{x^2(1-x)^2}
\int^{\infty}_{0}
\frac{d\nu\;\;\nu^2}{(\nu^2+\frac 14)^2}
\left(\frac{W^2}{Q^2+\mu^2_V}\right)^{\omega(\nu)}
\nonumber \\
&\times&
\int d^2\rho \; d^2R
e^{i\vec{\kappa}_\perp\left( \vec{R}-\left( \frac 12-x\right)
\vec{\rho}\right)}
\left[\frac{\rho^2}{\left(\vec{R}+\frac{\vec{\rho}}2\right)^2
\left(\vec{R}-\frac{\vec{\rho}}2\right)^2} \right]^{\frac 12 +i\nu}
F_{L,T}(x,\rho^2),
\end{eqnarray}
where $\vec{R}=(\vec{\rho}_1+\vec{\rho}_2)/2$ and
$\vec{\rho}=\vec{\rho}_1-\vec{\rho}_2$.
The functions $F_{L,T}(x,\rho^2)$ are defined as follows:
\begin{eqnarray}
F_{L}(x,\rho^2)&=&32\;x^2(1-x)^2\; \Phi^{(0)}_V(x,\rho^2)
\Phi^{(0)}_\gamma(x,\rho^2),
\\
F_{T}(x,\rho^2)&=&8\;\left( m^2 \Phi^{(0)}_V(\rho^2)
\Phi^{(0)}_\gamma(\rho^2)+
2\left( 1-2x(1-x) \right)
\Phi^{(1)}_V (x,\rho^2) \Phi^{(1)}_\gamma(x,\rho^2)\right),
\label{final}
\end{eqnarray}
where
\begin{eqnarray}
\Phi^{(n)}(\rho^2)&=&\int^{\infty}_{0}dk\;k^{n+1}\;\Psi(k^2) J_n(k\rho).
\label{deffi}
\end{eqnarray}

(ii) {\bf BFKL-pomeron-nucleon coupling}.
One could also imply another scenario for
the Pomeron coupling to nucleon
target. Indeed, if the gluon ladder is rather short and  BFKL Pomeron
is not transformed into a soft one, the amplitude of the vector meson
production takes the following form:
\begin{eqnarray}
A^{BFKL}_{L,T}(\gamma^* p\to Vp)&=&-i
\int^{1}_{0} \frac{dx}{x^2(1-x)^2}
\int^{\infty}_{0}
\frac{d\nu\;\;\nu^2}{(\nu^2+\frac 14)^2}
\left(\frac{W^2}{Q^2+\mu^2_V}\right)^{\omega(\nu)}
\nonumber \\
&\times&
\int d^2\rho \; d^2R
e^{i\vec{\kappa}_\perp\left( \vec{R}-\left( \frac 12-x\right)
\vec{\rho}\right)}
\left[\frac{\rho^2}{\left(\vec{R}+\frac{\vec{\rho}}2\right)^2
\left(\vec{R}-\frac{\vec{\rho}}2\right)^2} \right]^{\frac 12 +i\nu}
\;\;\frac 1{(2\pi)^2}\; C A_N(\nu,\vec{\kappa}_\perp)
S^{II}_{L,T}.
\label{finalnew}
\end{eqnarray}
Here we have introduced the quantity
\begin{eqnarray}
A_N(\nu,\vec{\kappa}_\perp)&=&
\int d^2\rho'_1 d^2\rho'_2
\left(
\left[\frac{(\vec{\rho'}_1-\vec{\rho'}_2)^2}{\rho'^2_1\rho'^2_2}\right]
^{\frac 12 -i\nu}
-\left[\frac 1{\rho'^2_1}\right]^{\frac 12 -i\nu}
-\left[\frac 1{\rho'^2_2}\right]^{\frac 12 -i\nu}
\right) \exp\left(i\vec{\kappa}_\perp \frac{\vec{\rho'}_1+\vec{\rho'}_2}
2\right)
\nonumber \\
&\times&
\left( \exp\left(-\frac{2(\rho'^2_1+\rho'^2_2)}{3<r^2>}\right)
-\delta(\vec{\rho'}_1-\vec{\rho'}_2)\frac 32\pi <r^2>
\exp\left(-\frac{(\vec{\rho'}_1+\vec{\rho'}_2)^2}{6<r^2>}\right)
\right)\ ,
\label{an}
\end{eqnarray}
with $<r^2>=0.8$ fm$^2$. This form of the nucleon wave function
corresponds to
the exponential dependence of the nucleon-pomeron vertex
$\exp(-B\kappa^2_\perp)$ with $B \simeq 2$ GeV$^{-2}$.

However, numerically the amplitude turns out to be poorly
sensitive
to the details of the mechanism of the pomeron attachment to the nucleon
target: namely, all the results obtained within
the assumiption of a direct
attachment of the BFKL pomeron to the nucleon coincide, within a
few percent accuracy, with the results obtained under the assumption
that the
transition from the BFKL regime to the soft one occurs before attaching to
the nucleon.

\subsubsection{Pomeron coupling to the quark loop}

Before, following refs. \cite{BFKL,Lip,ryskin,forshaw},
we have determined the $W$-dependence using the factor
$[W^2/(Q^2+\mu^2_V)]^{\omega (\nu )}$: here the variables $Q^2$ and
$\mu^2_V$ are external with respect to the quark loop. In writing spectral
representation for the quark loop, the more self-consistent procedure
is to use the internal variables, $M^2_{q\bar q}$ and
$M'^2_{q\bar q}$, instead of external ones.
For this scheme, we should
replace in the above formulas:
\begin{eqnarray}
\left (\frac{W^2}{Q^2+\mu^2_V}\right )^{\omega (\nu )} \to
\left (\frac{W^2}{M^2_{q\bar q}+M'^2_{q\bar q}}\right )^{\omega (\nu )}
\label{formfac}
\end{eqnarray}
The BFKL-equation with internal variables for the
dimensionless $W$-dependent factor was considered in \cite{Fadin}.
We perform calculations for both variants, using both external and
internal variables.

\subsubsection{Numerical calculations}

Numerical calculations have been performed using the
Monte-Carlo simulation program VEGAS \cite{veg}.
Integration limits related to $M^2_{q\bar q}$ and $M'^2_{q\bar q}$ depend
on $Q^2$, namely, $M^2_{q\bar q} \le 100\, Q^2$ (the increase of the upper
limit does not change the result). The integration over $\rho$ is
performed up to $\rho \le 5$ fm.

\section{Results}
To study the onset of the hard region dominance in diffractive processes
initiated by the photon with the increase of $Q^2$,
we consider in parallel the following reactions
\begin{eqnarray}
\gamma_L^*(Q^2)p\to\gamma_L^*(Q^2) p ,
\label{gamma1}
\end{eqnarray}
\begin{eqnarray}
\gamma_T^*(Q^2)p\to\gamma_T^*(Q^2) p
\label{gamma2}
\end{eqnarray}
and
\begin{eqnarray}
\gamma_L^*(Q^2)p\to\rho_L^0 p ,
\label{rho1}
\end{eqnarray}
\begin{eqnarray}
\gamma_T^*(Q^2)p\to\rho_T^0 p .
\label{rho2}
\end{eqnarray}
The main ingredients of the calculation of the amplitudes of these reactions
are the consideration of the hard quark block and the pomeron amplitude and
its attachment to the nucleon target.

The processes (\ref{gamma1}), (\ref{gamma2})
can be reliably described
at large $Q^2$, because the photon wave function is known
due to the analysis of ref. \cite{AMN_a}. The corresponding vertex
$G_{\gamma}(M^2_{q\bar q})$ is shown in Fig. 3a. Notice that a reliability of
calculating the quark loop diagram
$\gamma^*(Q^2) \to q\bar q \to \gamma^*(Q^2)$
is the main motivation for the study of
the processes (\ref{gamma1}),
(\ref{gamma2}). The calculated cross section
of the reaction (\ref{gamma1}) within the
amplitude determined by Eq. (49)
is shown in Fig. 4 by solid lines.
The amplitude $A_L$ determines $d\sigma
/d\kappa_{\perp}^2
(\gamma_L^*(Q^2)p\to\gamma_L^*(Q^2) p) $ (Figs. 4b,c,d): the
integrated $\kappa_{\perp}^2$-distribution
over the region
$\kappa_{\perp \; min}^2 \leq \kappa_{\perp}^2 \leq 1$ GeV$^2$
yields $\sigma (\gamma_L^*(Q^2)p\to\gamma_L^*(Q^2) p)$
(the cutting parameter $\kappa_{\perp \; min}^2=0.05$ GeV$^2$
is introduced to avoid the divergence of the BFKL-amplitude at
$\kappa_{\perp }^2=0$).

For illustration of the role of low-$M_{q\bar q}$ region in the
realistic photon vertex, $ G_{\gamma}(M^2_{q\bar q})$, we also perform
calculations of the cross section $\gamma_L^*(Q^2)p\to\gamma_L^*(Q^2) p$
with  $G_{\gamma}(M^2_{q\bar q})=1$, Fig. 5.

Results for the reaction (\ref{gamma1}) are shown in Fig. 6.

An essential ingredient of the amplitude of the reaction is the
description of the pomeron-exchange block.
A realistic picture of the pomeron block is the following:
small average transverse separations in the gluon ladder selected by the
quark loop increase as a result of the $t$-channel evolution, and the hard
pomeron transforms into a soft one along the gluon ladder.
This feature is taken into account by using
the pomeron-proton coupling as given in Eq. (49).
The corresponding results are shown in Figs. 4-8.

Nevertheless, for the sake of illustration we also performed
calculations for another variant, with a
prompt attachment of the BFKL-pomeron directly to the target nucleon,
see Eq. (\ref{an}). The results for the $W$-dependence,
after renormalizing the coupling constant $C=g^2\tilde g$, are
practically the same for both variants so we do not
present separately the results obtained by using Eq. (\ref{an}).

We also study different possibilities of choosing the scale of the
$W^2$-dependence of the BFKL amplitude and consider the two
possibilities:
(1) $W^2/(Q^2+\mu^2_V)$ and
(2) $W^2/(M^2_{q\bar q}+M'^2_{q\bar q})$, see
Eq. (\ref{formfac}).
Calcuations with the variant (2)  are shown in
Fig. 7. One observes a sizeable sensitivity of the
calculated cross sections to the choice of the $W^2$-scale.

The reactions of the $\rho^0$-meson diffractive production
are extensively studied both experimentally and theoretically
(see, e.g., \cite{forshaw,martin,data} and references therein).
However, the calculation of the quark-loop diagram
$\gamma^*(Q^2) \to q\bar q \to \rho^0$
turns out to be rather ambiguous due to the uncertainty in the
large-$M_{q\bar q}$ behaviour of the $\rho$-meson vertex
$G_{\rho}(M^2_{q\bar q})$.
Because of that, we analyze several possibilities of the
large-$M_{q\bar q}$ behaviour of the vertex $G_{\rho}(M^2_{q\bar q}) $.
The results for the two
variants, namely (1) $ G_{\rho}(M^2_{q\bar q}) \sim M^{-2}_{q\bar q}$ and
(2) $G_{\rho}(M^2_{q\bar q}) \sim \; const $  (see Fig. 3b), are represented
by solid lines in Fig. 8.

In order to check the consistency of our initial assumption that at
the considered $Q^2$ and $W$ the ampitude is dominated by the small
separations of the gluons in the ladder, and thus the BFKL form of the kernel
should be used we have also performed calculations
introducing explicitly a $\theta$-function cut
into the BFKL-amplitude integrand of (49): this $\theta$-function
measures an actual contribution of the hard region of separations smaller
than 0.2 fm:
\begin{eqnarray}
d^2\rho \to d^2\rho\; \theta (0.2\;fm- \rho).
\label{0.2fm}
\end{eqnarray}
The cross sections constrained by (\ref{0.2fm}) are shown
in Figs. 4, 5, 6, 7 and 8 by dashed lines:
one can see that for all reactions
the selection of small distances, if any, occurs very slowly with the
increase of $Q^2$. Only at $ Q^2>100 $ GeV$^2 $, the 80\% of the cross sections
are actually gained in the hard region. Thus we conclude that
the dominance of the hard region cannot be expected earlier than at
$ Q^2>50 - 100 $ GeV$^2 $.

Let us notice that the situation with the diffractive production
by the hard photon in the region
of $Q^2\le$ 100 GeV$^2$ turns out
to be quite similar to the situation with the elastic meson form factor
at $Q^2 \le$ 10-20 GeV$^2$:
in the latter case one can expect the dominance of the hard-scattering
mechanism (which is proved to be the dominant mechanism at
asymptotically large $Q^2$) also at $Q^2$ of several GeV$^2$.
Assuming such dominance at several $Q^2$
one determines the soft wave function of the pion at low normalization point
by descrining the data on the form factor and finds for this wave function a
double-humped form \cite{ZhO}. However, analyzing the content of the
calculated form factor, one finds that the bulk of the contribution actually
comes from the end-point region where the hard-scattering mechanism is not
applicable but rather the Feynman mechanism works \cite{LSI}. This analysis
shows that the pertutbative treatment of the form factor at several GeV$^2$
is not consistent (or at least the perturbative mechanism cannot give a bulk
of the form factor) and that the soft physics actually dominates in the
kinematical region $Q^2\le 20$ GeV$^2$.

Likewise, we have found that the assumption of the BFKL form of the pomeron
kernel actually yields in the region of $Q^2\le 100$ GeV$^2$ the cross
section which is at 80\% level gained in the region
of {\it large} transverse separations in the quark loop.
The latter are equal to gluon separations at the top of the gluon ladder
which thus turns out to be in the soft regime just from the top.
Therefore, similar to the elastic form factor case, we have to conclude
that the perturbative treatment in this range of $Q^2$ is not consistent and
that rather the soft pomeron should be used.

Considering the region of $Q^2\ge$ 100 GeV$^2$, we have observed that the
dominance of the region of small separations in the quark loop depends on the
subtle details of the calculation procedure: namely, some of the variants of
the calculation discussed here yield the cross sections $\sigma
(W,Q^2)$ and $(\sigma (W,Q^2))_{\rho < 0.2\; fm}$ which differ even at
very large $Q^2$: this means that the region  $\rho > 0.2$ fm still
gives a non-vanishing contribution even at asymptotically large $Q^2$.

More detailed and technical explanation is as follows:

It might happen that the spectral representation of the quark loop was
superconvergent, i.e. the factor $M^2_{q \bar q}$ in the denominator
was not essential for the convergence of the integral.  Then the
denominator $(M^2_{q \bar q} +Q^2)^{-1}$ could be safely expanded in
powers of $1/Q^2$, namely, $(M^2_{q \bar q} +Q^2)^{-1} \to 1/Q^{2}$.
Then one would not observe any selection of the hard region even at
asymptotically large $Q^2$: the $Q^2$ dependence would just factorize
and the integrals would be still dominated by typical hadronic scales.

In the reactions (\ref{gamma1}) and
(\ref{gamma2}), choosing the scale $W^2/(Q^2+\mu^2_V)$ for the
$W^2$-dependence, one observes the cross sections $\sigma(W,Q^2)$ and
$(\sigma (W,Q^2))_{\rho < 0.2\; fm}$
to be very close to each other at $Q^2 \ge 100$ GeV$^2$ (see Figs.
4 and 6). This means that the spectral representations are not
superconvergent and thus the hard photon actually selects small distances in
the quark loop and, as a result, the top of the pomeron ladder is in the
perturbative regime.

The introduction of the scale-factor
$W^2/(M^2_{q\bar q}+M'^2_{q q})$ provides a superconvergence of the spectral
integrals, thus making the cross sections
$\sigma (W,Q^2)$ and  $(\sigma (W,Q^2))_{\rho < 0.2\; fm}$
different even at $Q^2 \to \infty$. In other words, small separations are not
selected in this variant of calculation even at asymptotically large $Q^2$.

The similar situation is observed in the diffractive vector meson
production reactions (\ref{rho1}) and (\ref{rho2}).

Namely, for the variant $G_{\rho}(M^2_{qq}) \sim Const$ at
$M^2_{qq} \to \infty$ (see Fig. 8) the integrals are not superconvergent and
thus small separations dominate the amplitude at large $Q^2$.  The only
quantitative difference with the photon production case is that the
proximity of cross sections $\sigma (W,Q^2)$ and $(\sigma (W,Q^2))_{\rho <
0.2\; fm}$, with the scale $W^2/(Q^2+\mu^2_V)$, comes later with the
increase of $Q^2$.

For the variant $G_\rho(M^2_{q\bar q}) \sim 1/M^2_{q\bar q}$
the spectral representations are superconvergent and there is no dominance of
the small separations in the quark loop even at asymptotically large $Q^2$;
correspondingly, the contribution of distances $\rho > 0.2$ fm is always
important in this variant.

Low-$M_{q\bar q}$ structure of
$G_\rho (M^2_{q\bar q})$ and $G_\gamma (M^2_{q\bar q})$ is very
important in realistic treatment of reactions
$\gamma^*(Q^2)p\to V p$ and $\gamma^*(Q^2)p\to \gamma^*(Q^2) p$:
it is seen in comparing   Figs. 4 and 5.
Low-$M_{q\bar q}$ structure is essential for different behaviour
of  $d\sigma /d\kappa ^2_{\perp}(W,Q^2,\kappa ^2_{\perp})$ and
$(d\sigma /d\kappa ^2_{\perp}(W,Q^2,
\kappa ^2_{\perp}))_{\rho < 0.2\; fm}$ in the region of small
$\kappa ^2_{\perp}$ at $Q^2 \sim 20 - 50$ GeV$^2$.

Let us discuss now in a more detail the role of large and small
$M^2_{q\bar q}$ in the formation of cross sections initiated by
$\gamma^*_L(Q^2)$ and $\gamma^*_T(Q^2)$. The distinction of
corresponding cross sections is due to a different structure of the
spin-dependent factors, $S^{II}_T$ and $S^{II}_L$, related to the loop
diagrams (see (\ref{ssp2}) and (\ref{sssp2}): for the case
of longitudinal polarization the spin dependent factor is
$S^{II}_L \sim x^2(1-x^2)$, whereas $S^{II}_T \ne 0$ at $x \to 0$ or
$(1-x) \to 0$. Therefore, since $M^2_{q\bar q}=m^2_\perp/x/(1-x)$, the
large masses, i.e. the regions $x \sim 0$ and $(1-x) \sim 0$, become
dominant for the reactions $\gamma^*_T(Q^2) \to \gamma^*_\pi (Q^2)$ and
$\gamma^*_T(Q^2) \to \rho^0_T $, but there is no such dominance
for the reactions with $\gamma^*_L(Q^2)$. This means that in the
transitions $\gamma^*_L(Q^2) \to \gamma^*_\pi (Q^2)$
$\gamma^*_L(Q^2) \to \rho^0_L $ small $M^2_{q\bar q}$ or large
interquark separations contribute considerably. The realistic photon
wave function also enhances the role of the large interquark distances,
thus resulting in a different behaviour of the cross sections
$d\sigma/d\kappa_\perp^2\left(\gamma^*_L(Q^2)p \to \gamma^*_L(Q^2)p
\right )$ and
$d\sigma/d\kappa_\perp^2\left(\gamma^*_L(Q^2)p \to \gamma^*_L(Q^2)p
\right )_{\rho < 0.2\,{\rm fm}} $ at $\kappa^2_\perp 0.1$ GeV$^2$ in
the region $Q^2 \sim 20-50$ GeV$^2$ (compare Fig. 4c and Fig. 5d).

\section{Conclusion}

The performed calculations demonstrate that a selection of small
distances by the virtual photon $\gamma^*(Q^2)$ with the increase of
$Q^2$ have the following features:
First, it strongly depends on the choice of the $W^2$-scale.
Second, it proceeds very slowly:  in the quark loops  for the transitions
$\gamma^*(Q^2) \to \gamma^*(Q^2) $ and
$\gamma^*(Q^2)\to\rho^0 $ the soft interquark distances $ \rho >0.2$ fm
clearly dominate the amplitude at $Q^2<100$ GeV$^2$.

The situation here is quite similar to that of meson elastic form factors,
where the pQCD regime starts to dominate at $Q^2>50$ GeV$^2$ only
(see the results for the pion and transition form factors in \cite{AMN,AMN_a}
and a general discussion in \cite{LSI}). From this point of view, the results
of our analysis \cite{AMN,AMN_a} and the present paper seem to be quite
consistent with each other.

The late onset of the pQCD in the reaction
$\gamma^*(Q^2)p \to Vp $ means that it is the Strong-QCD pomeron that actually
works at $Q^2 \leq 100$ GeV$^2$, thus exposing an intriguing
situation with vector meson electroproduction processes.

The matter is that the $W$-dependence of the photoproduction
 reactions ($Q^2=0$)
in the region $W\sim 10-200$ GeV is rather flat, similar to that
in hadronic processes like $\pi p\to \pi p$. However, the reaction
$\gamma^*(Q^2)p \to Vp $  at $Q^2 \sim  10-20$ GeV$^2$  demonstrates
a different $W$-dependence: namely, the cross sections increase
as $W^{2\Delta}$ with $\Delta \approx 0.3$.
There might be an attractive possibility to refer this growth to the change of
the pomeron regime from the Strong-QCD one in the first case to the BFKL
pomeron in the second one.

However the results of our analysis show that the BFKL pomeron cannot be
seen in the diffractive production in the region $Q^2\le$ 100 GeV$^2$.
Therefore, a theoretical explanation of the change of the $W$ dependence
lies in better understanding of properties of the Strong-QCD pomeron.
One of the possibilities is to explain
this change of the $W$ dependence by a complicated
'heterogeneous' structure of the Strong-QCD pomeron, when its different
components reveal themselves at $Q^2\sim 0$ and $Q^2\sim 10$
GeV$^2$.

For example, one may suggest the following scenarios for the Strong-QCD
pomeron:
\begin{itemize}
\item
In the reactions with $Q^2 \sim 0$ as well as hadronic
processes $\pi p \to \pi p$ or $p p \to p p$, the amplitude is
governed by the pomeron with intercept close to unity: $j=1+\Delta$,
with $\Delta \simeq 0.1$ (KTDL-pomeron \cite{KTDL}). In this case,
the KTDL-pomeron contribution must vanish at $Q^2 \sim 10$
GeV$^2$ leaving the leading role to a new pole with a larger intercept,
$\Delta_{new\; pole} \simeq 0.3$ \cite{L}.
\item
In the reactions with $Q^2 \sim 0$ as well as  hadronic
processes $\pi p \to \pi p$ or $p p \to p p$, the amplitude is
determined by multiple primary-pomeron-induced rescatterings in the
direct channel \cite{eikonal,mila}. A slow growth of the
amplitudes at $W\sim 10-200$ GeV, $\Delta_{effective} \simeq 0.1$,
occurs at rather large value of the primary pomeron intercept,
$\Delta_{primary\; pomeron}\simeq 0.3$ \cite{mila}. In this
scenario, in order to obtain the growth of the electroproduction
amplitudes observed in the experiment, $\sim W^{0.6}$,
the multiple rescatterings
should die with the increase of $Q^2$ in such a way that at moderately
large $Q^2$ only the one-pomeron exchange survives, leading to the amplitude
growth $\sim W^{2\Delta_{primary\,pomeron}}$ .
\end{itemize}

However a detailed consideration of the Strong-QCD pomeron
is beyond the scope of this article and we leave this
intriguing subject to other publications.

\section{Acknowledgments}
We are grateful to G. Korchemsky for helpful comments.
This work was supported in part by RFBR under grant 98-02-17236.

\begin{figure}
%fig.1
\centerline{\epsfig{file=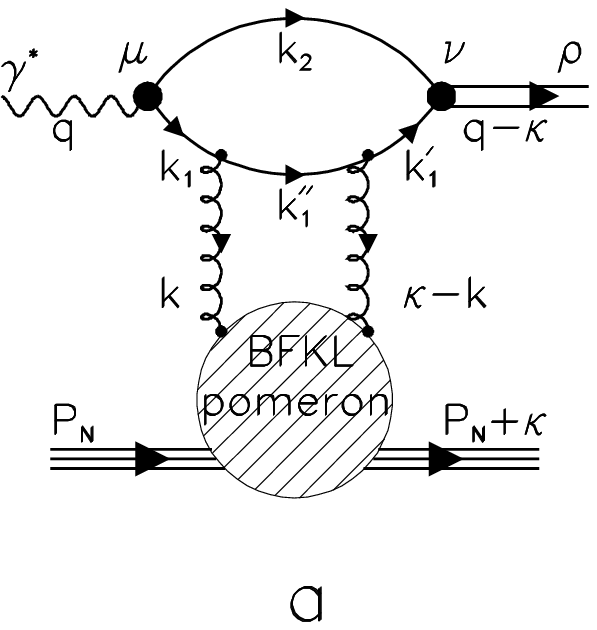,width=5.0cm}\hspace{3cm}
            \epsfig{file=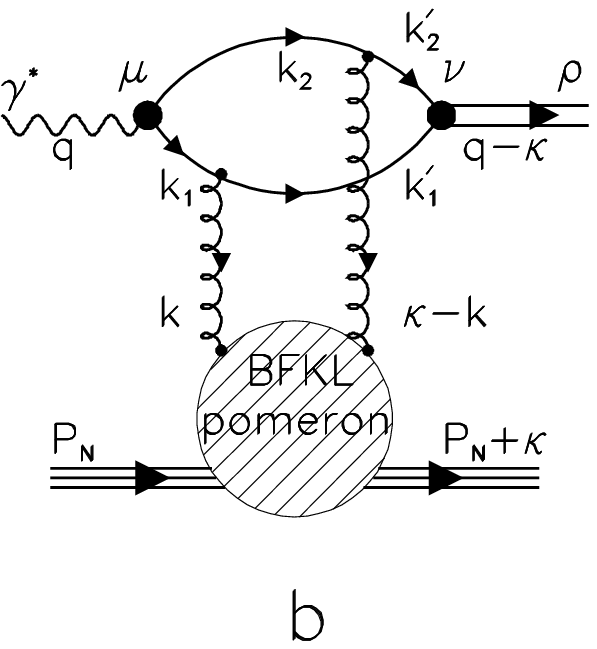,width=5.0cm}}
\vspace{1cm}
\centerline{\epsfig{file=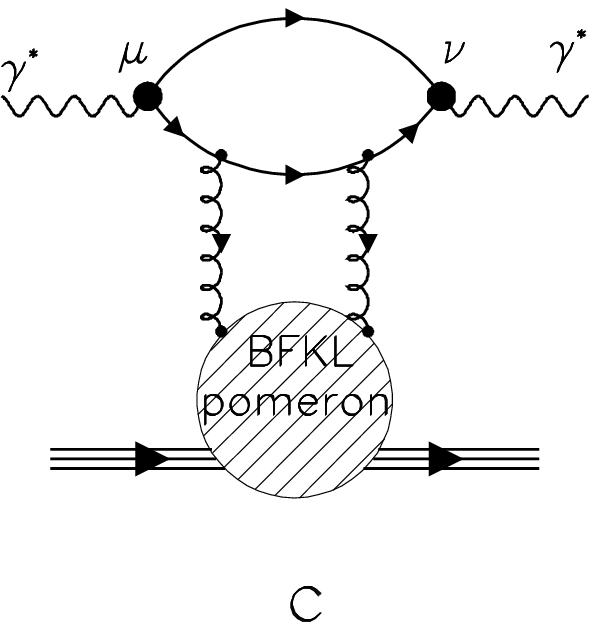,width=5.0cm}\hspace{3cm}
            \epsfig{file=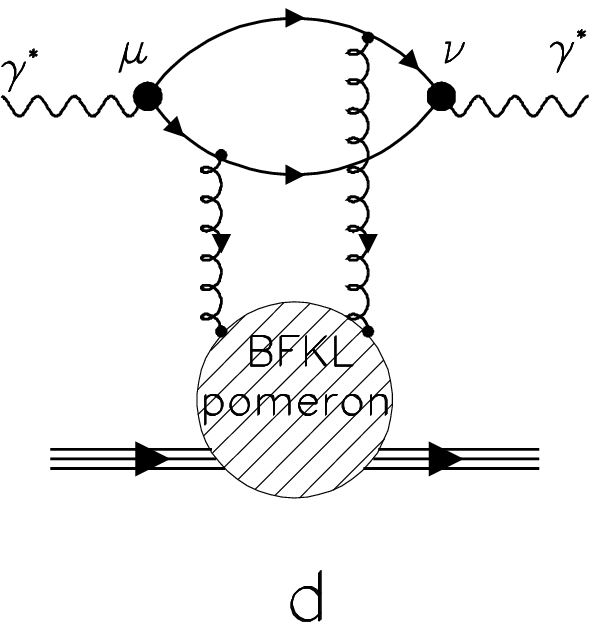,width=5.0cm}}
\vspace{1cm}
\caption{Diagrams for the processes $\gamma^*(Q^2)p\to\rho^0p$ (a,b)
and $\gamma^*(Q^2)p\to \gamma^*(Q^2) p$ (c,d).}
\end{figure}

\begin{figure}
%fig.2
\centerline{\epsfig{file=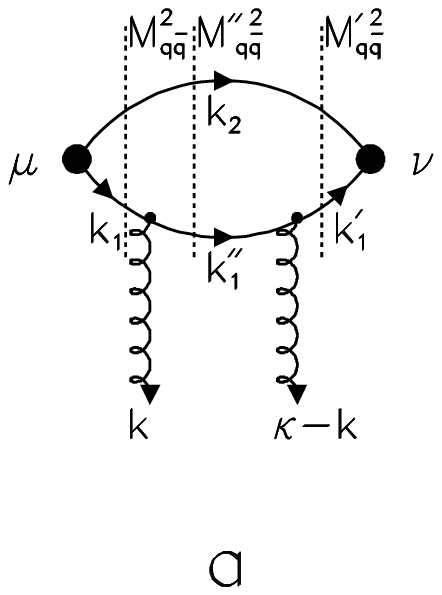,width=5.0cm}\hspace{3cm}
            \epsfig{file=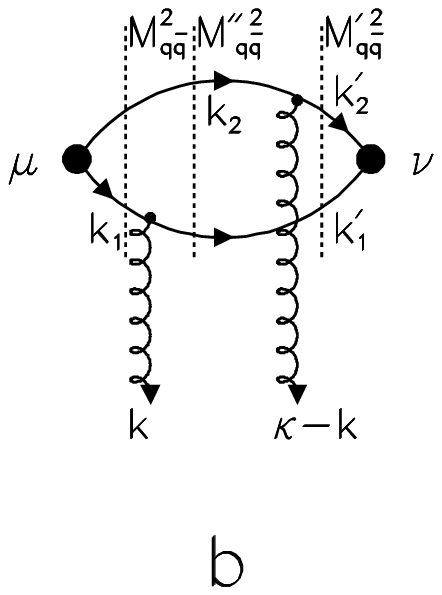,width=5.0cm}}
\vspace{1cm}
\caption{Quark loop for the case the gluon interaction with a quark
(a) and with quark and antiquark (b). Dashed line indicates the cut of
a loop related to the spectral integrals.}
\end{figure}

\begin{figure}
%fig.3
\centerline{\epsfig{file=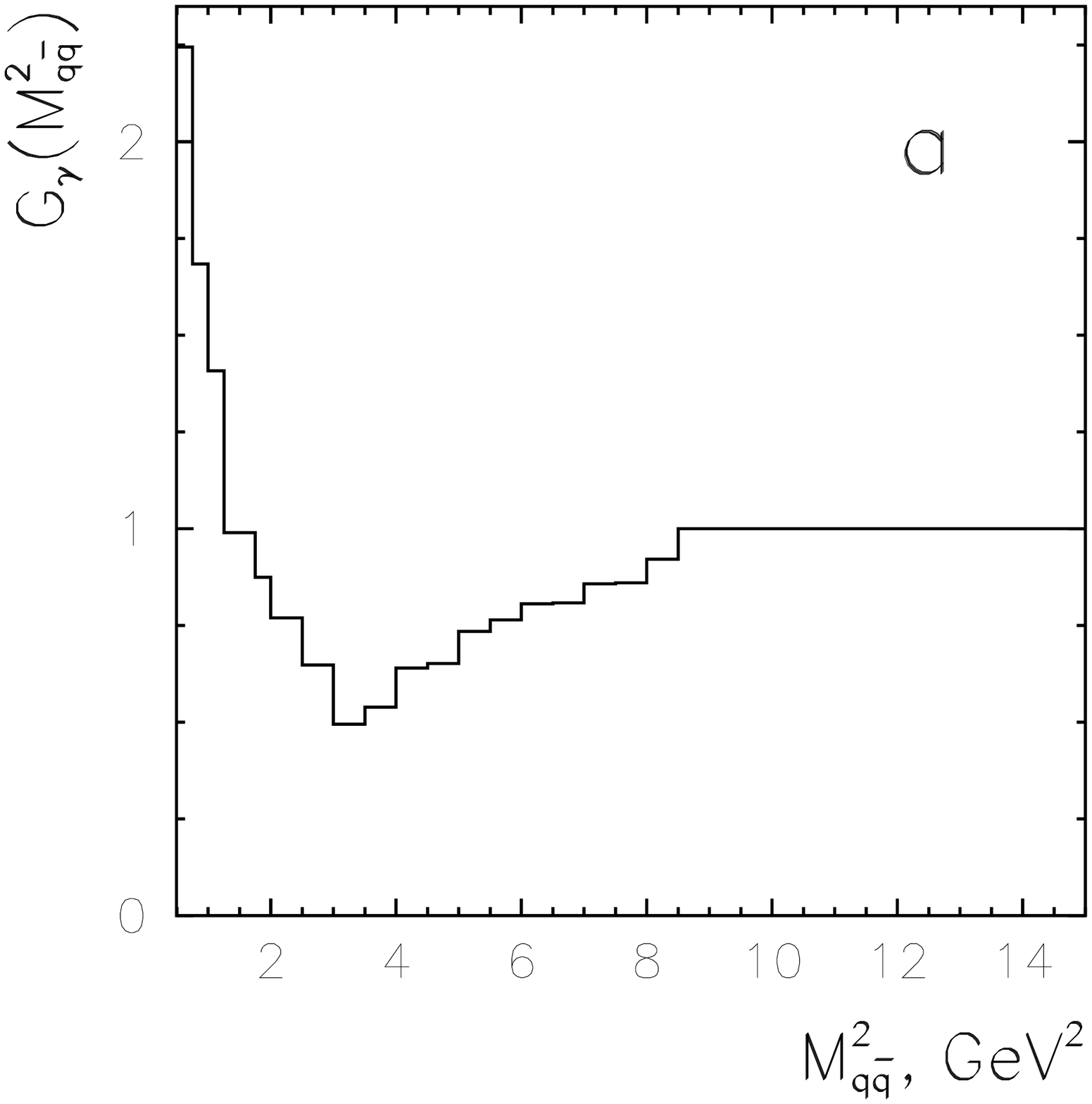,width=8.0cm}\hspace{1cm}
            \epsfig{file=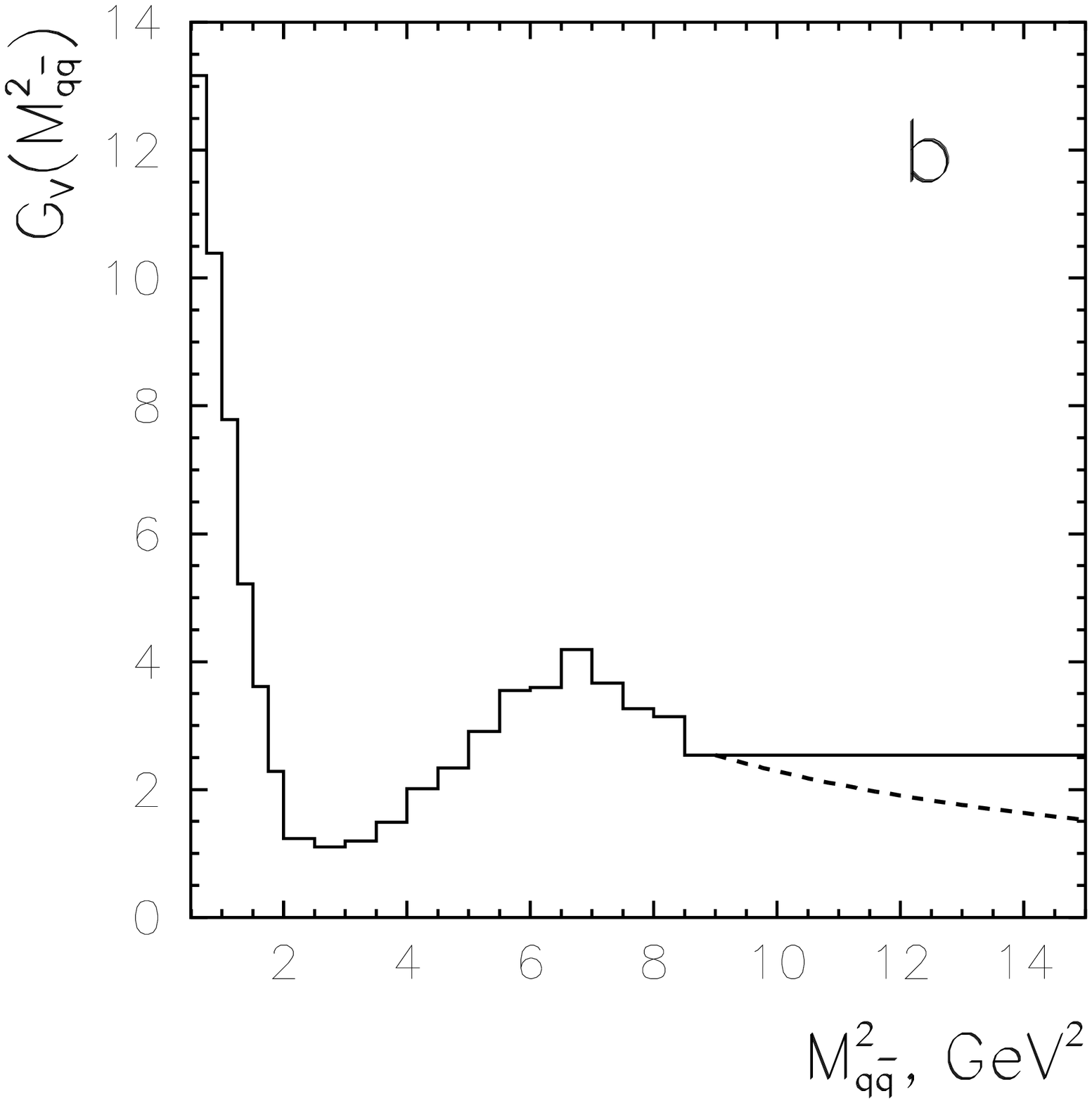,width=8.0cm}}
\vspace{1cm}
\caption{(a) The vertex $G_\gamma(M_{q\bar q}^2)$ found in the
analysis of transition form factors $\gamma\to\pi^0,\eta,\eta'$ in
[2]. (b) The $\rho$-meson vertex function used in the
calculation; At $M_{q\bar q} > 3$ GeV two variants are shown: I) the
decrease $G_\gamma(M_{q\bar q}^2) \sim 1/M_{q\bar q}^2$ (dashed
curve), II) $G_\gamma(M_{q\bar q}^2) \sim const$ (solid curve).}
\end{figure}

\begin{figure}
%fig.4
\centerline{
\epsfig{file=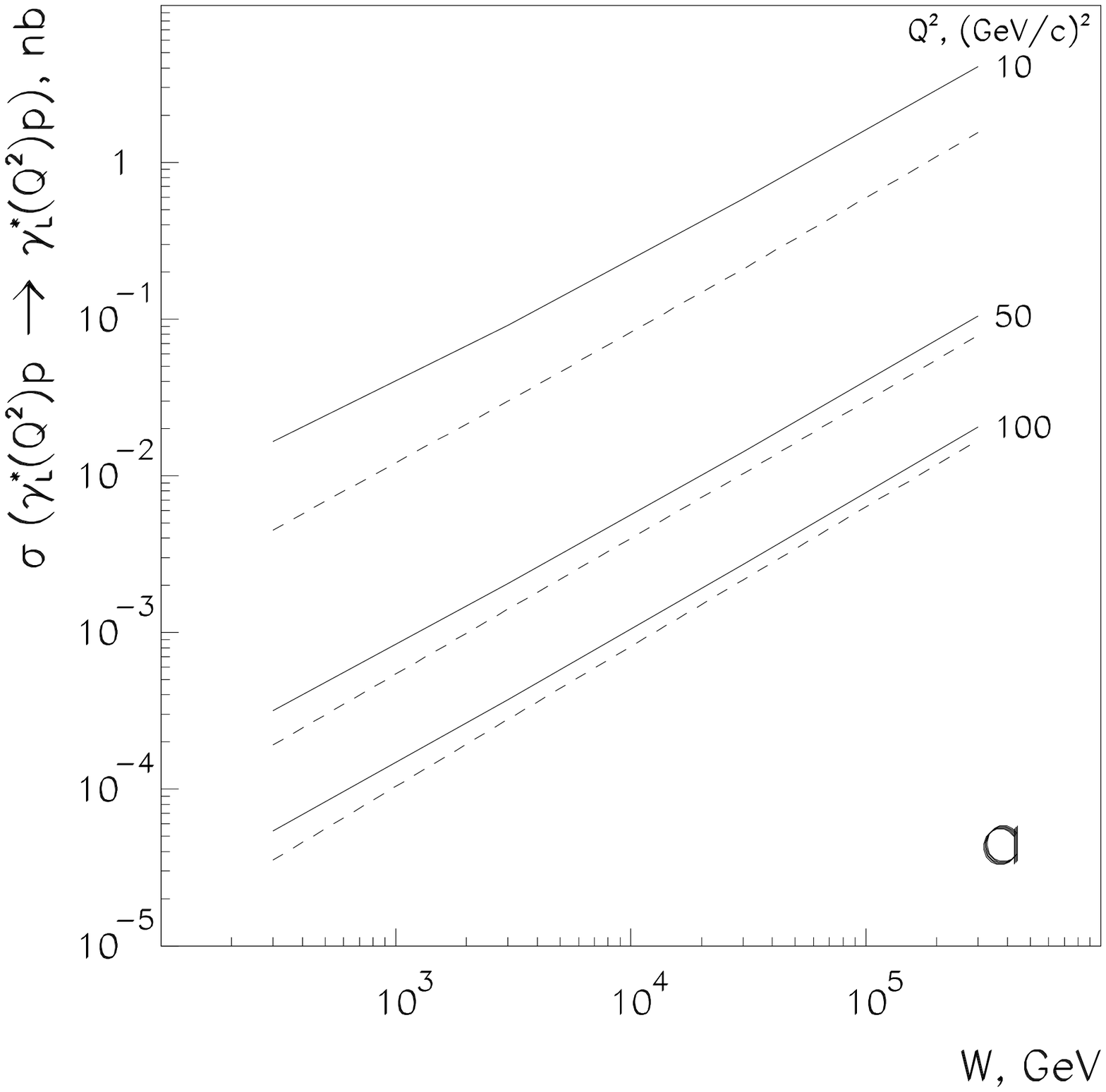,width=8.0cm}\hspace{1cm}
\epsfig{file=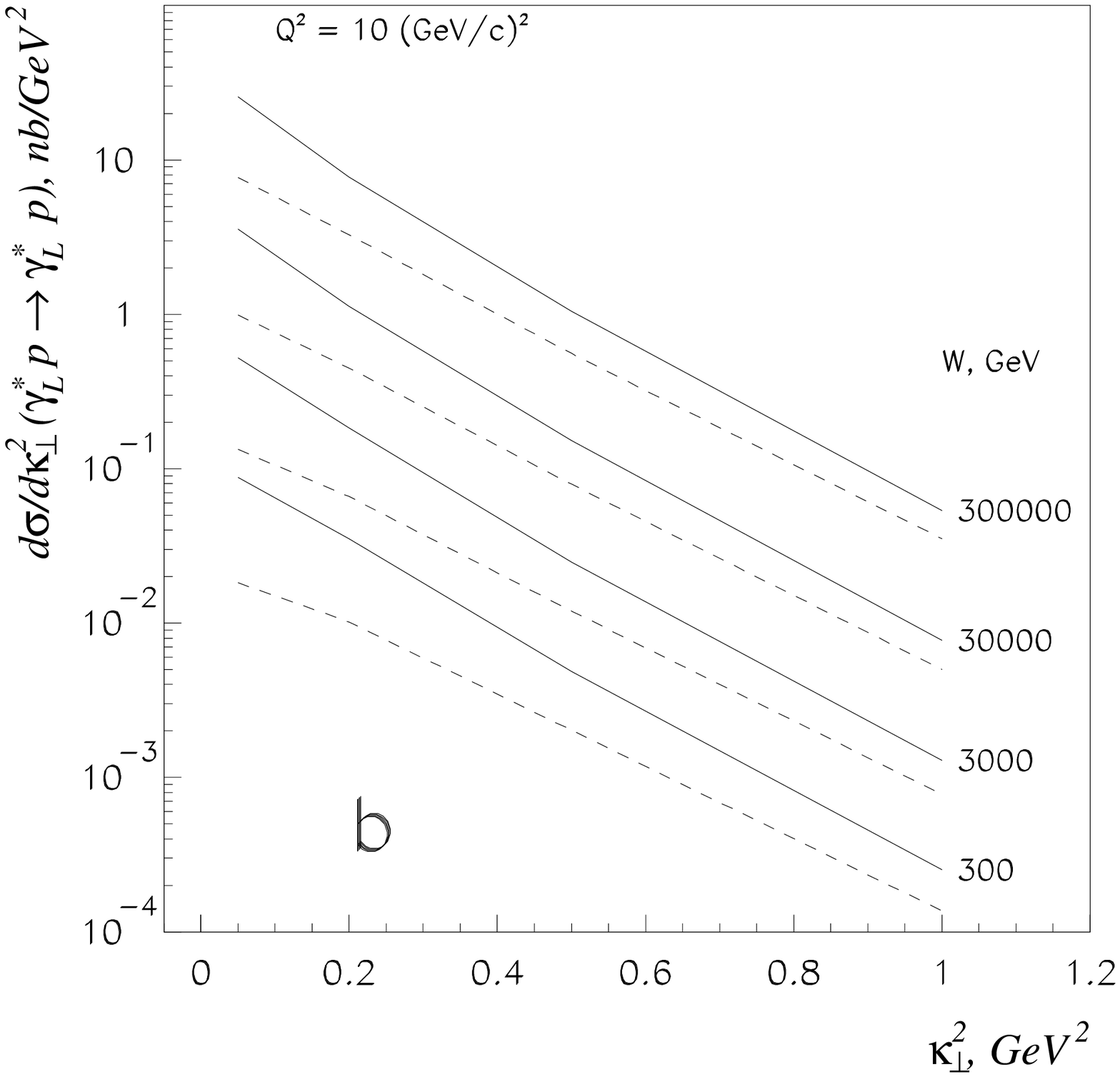,width=8.0cm}}
\vspace{1cm}
\centerline{
\epsfig{file=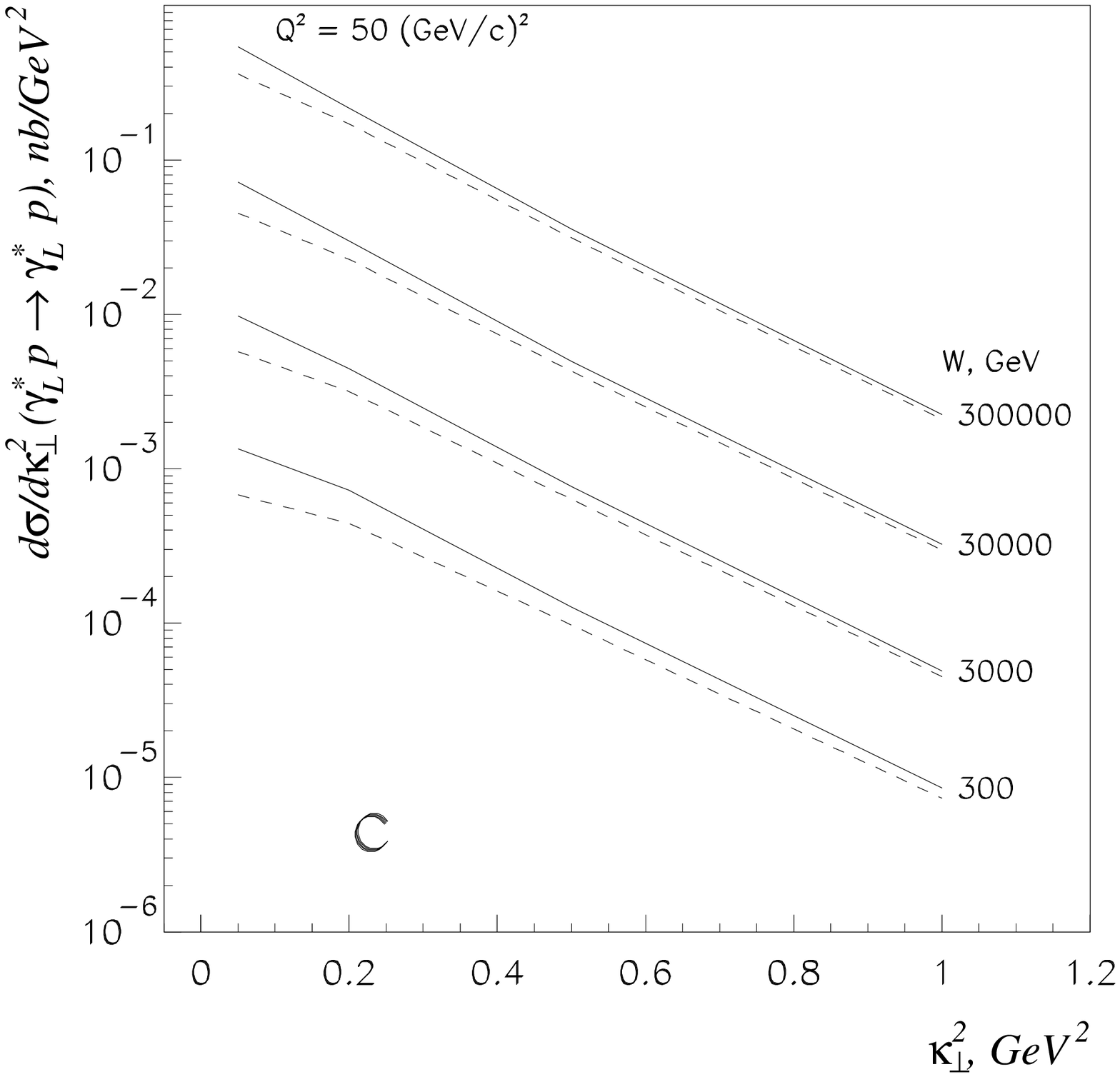,width=8.0cm}\hspace{1cm}
\epsfig{file=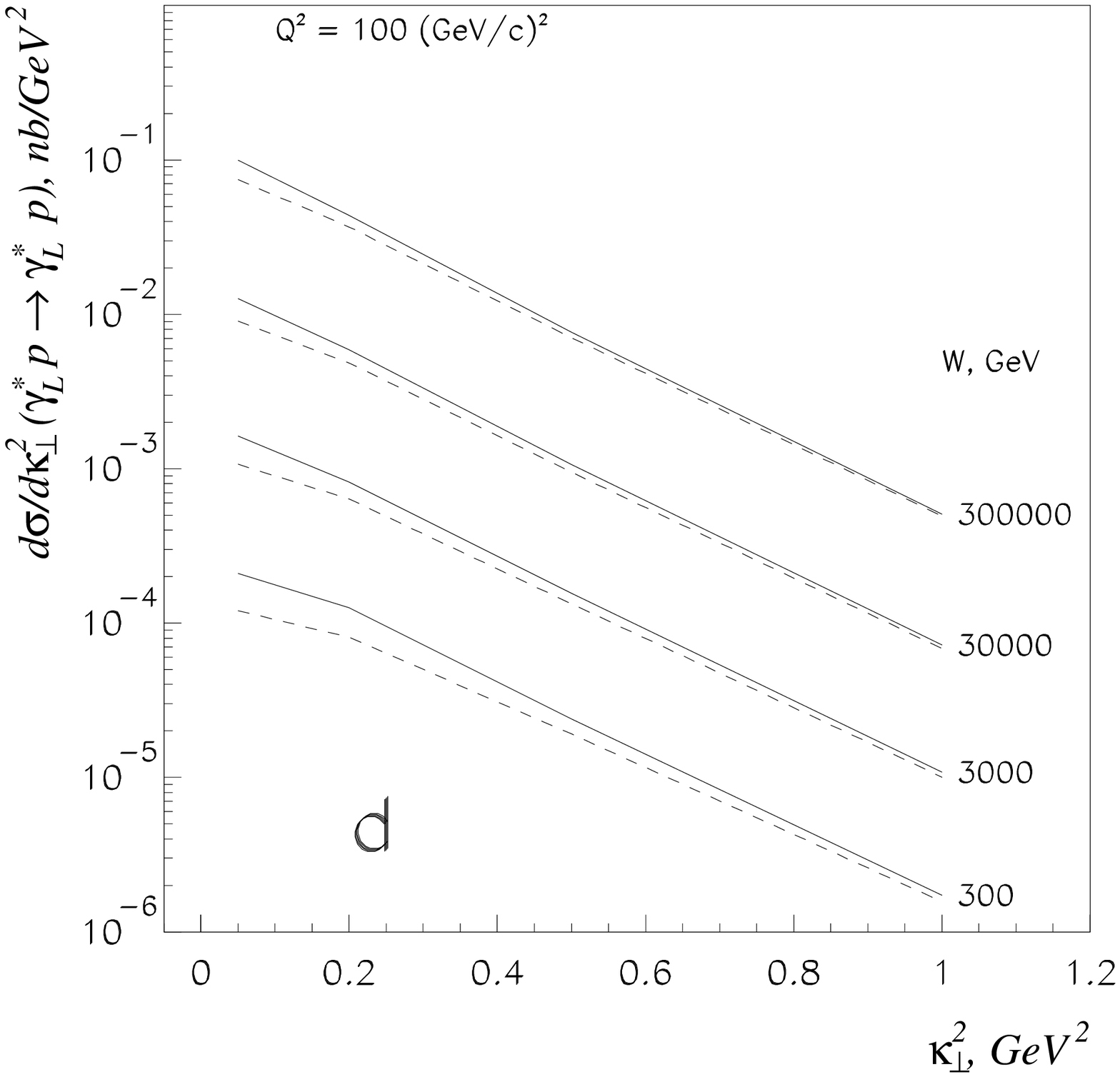,width=8.0cm}}
\vspace{1cm}
\caption{Cross sections in reaction
$\gamma^*_L(Q^2)p\to\gamma^*_L(Q^2)p$
with the $W$-scale factor $W^2/(Q^2+\mu^2_V)$, normalization is
fixed by cross section of reaction
$\gamma^*_L(Q^2=10\; GeV^2)p\to\rho^0_Lp$.
Solid curves correspond to
$\sigma(W,Q^2)$ (a) and $d\sigma/d\kappa^2_\perp(W,Q^2)$ (b,c,d)
calculated without a cut in the quark loop; dashed lines stand for
$\sigma(W,Q^2)_{\rho < 0.2\; fm}$ (a) and
$d\sigma/d\kappa^2_\perp(W,Q^2)_{\rho < 0.2\; fm}$ (b,c,d).}
\end{figure}

\begin{figure}
%fig.5
\centerline{
\epsfig{file=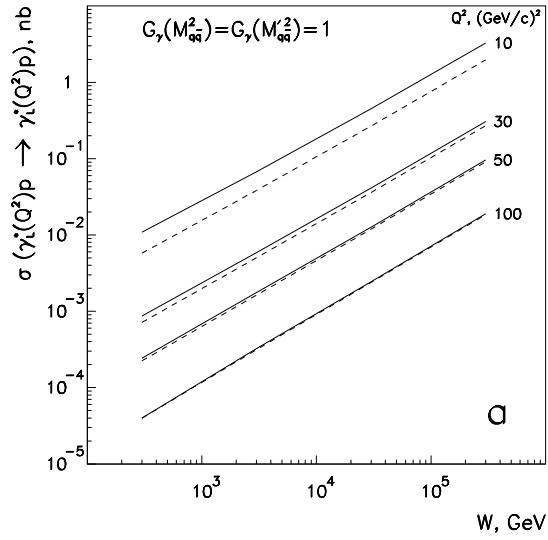,width=8.0cm}\hspace{1cm}
\epsfig{file=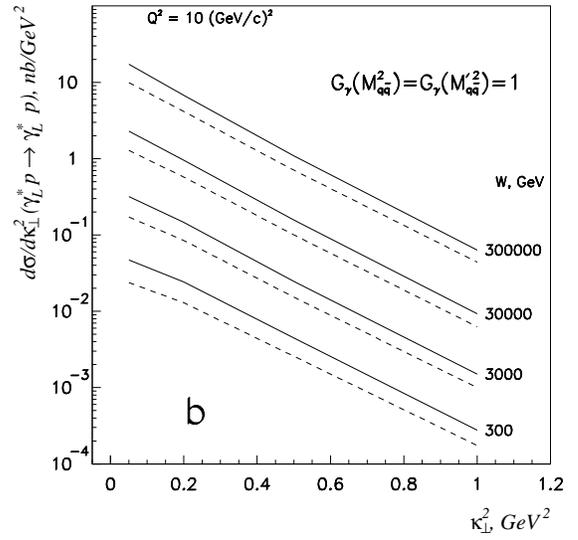,width=8.0cm}}
\vspace{1cm}
\centerline{
\epsfig{file=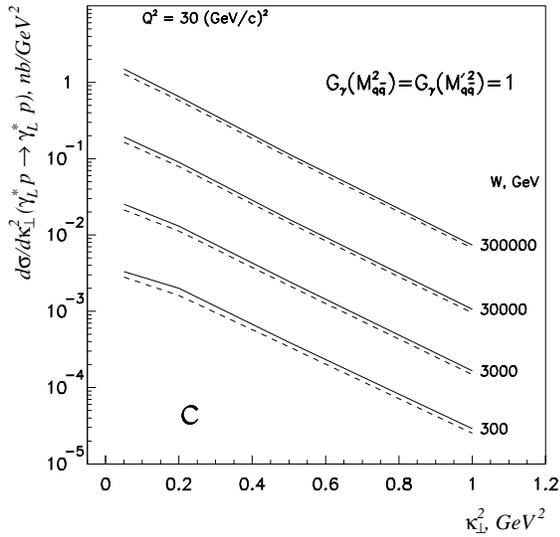,width=8.0cm}\hspace{1cm}
\epsfig{file=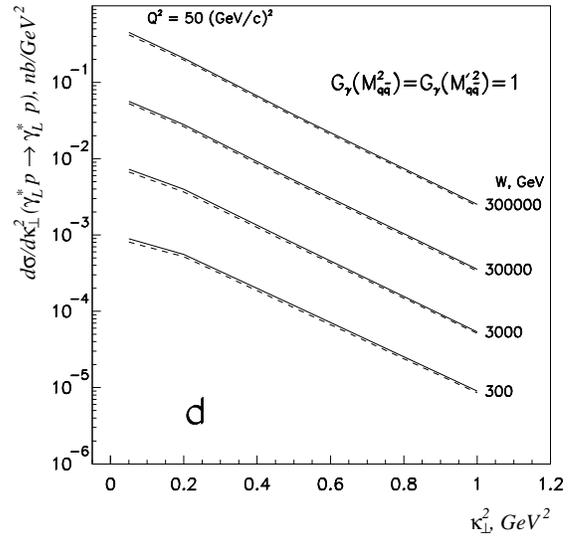,width=8.0cm}}
\caption{The same as in Fig. 4, but with
$G_\gamma(M^2_{q\bar q})=G_\gamma(M'^2_{q\bar q})=1$.}
\end{figure}

\begin{figure}
%fig.6
\centerline{
\epsfig{file=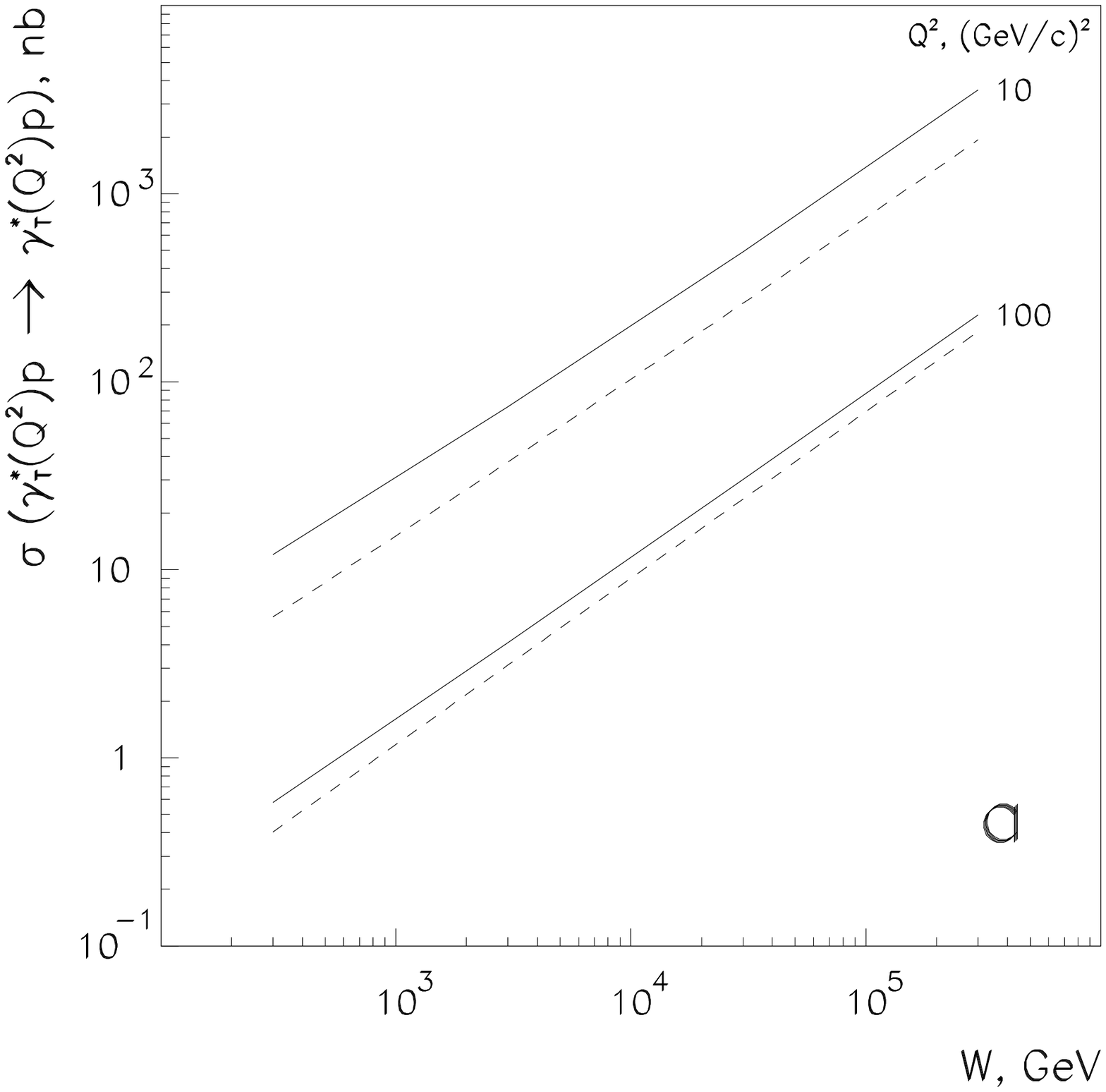,width=8.0cm}\hspace{1cm}
\epsfig{file=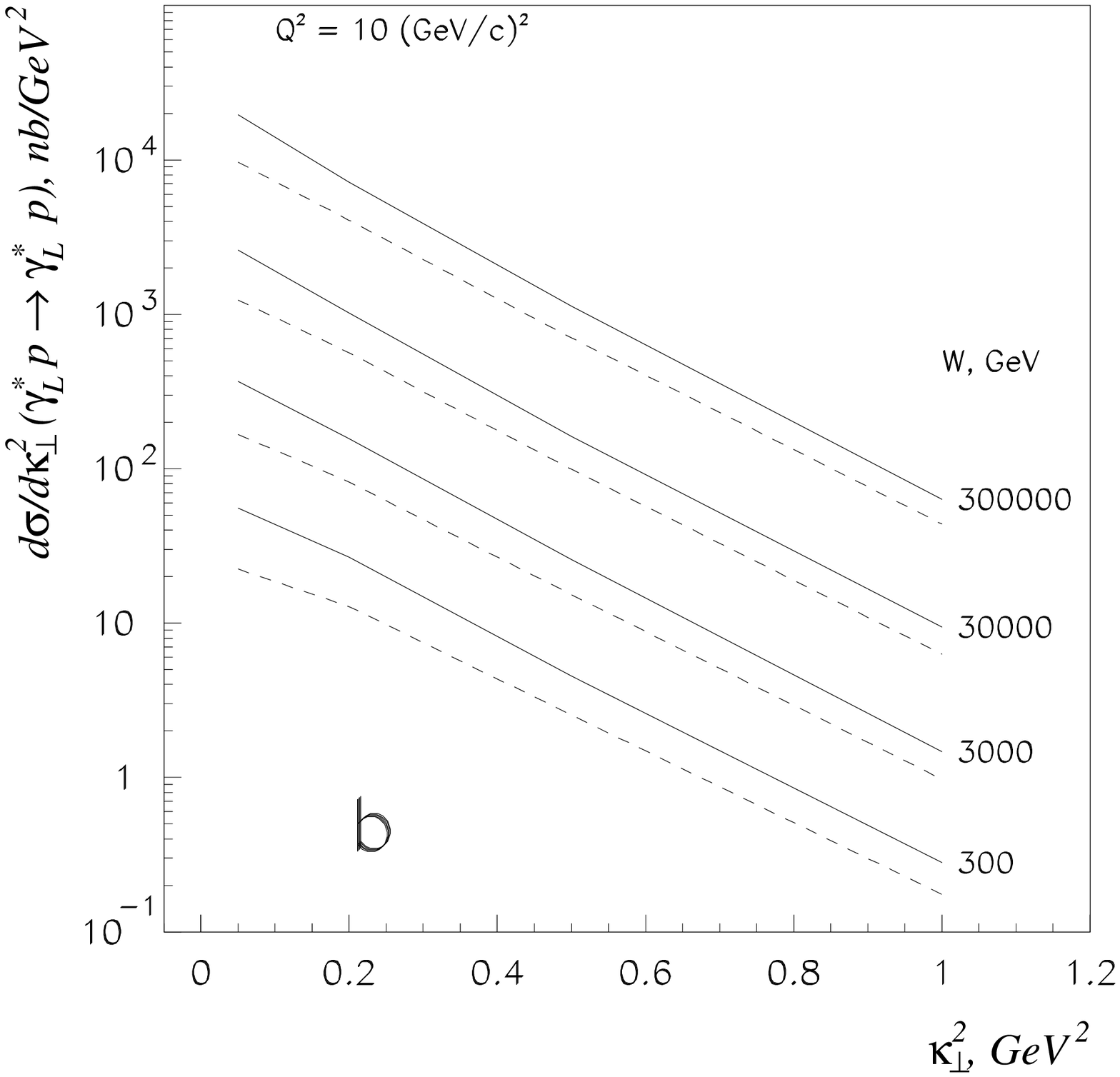,width=8.0cm}}
\vspace{1cm}
\centerline{
\epsfig{file=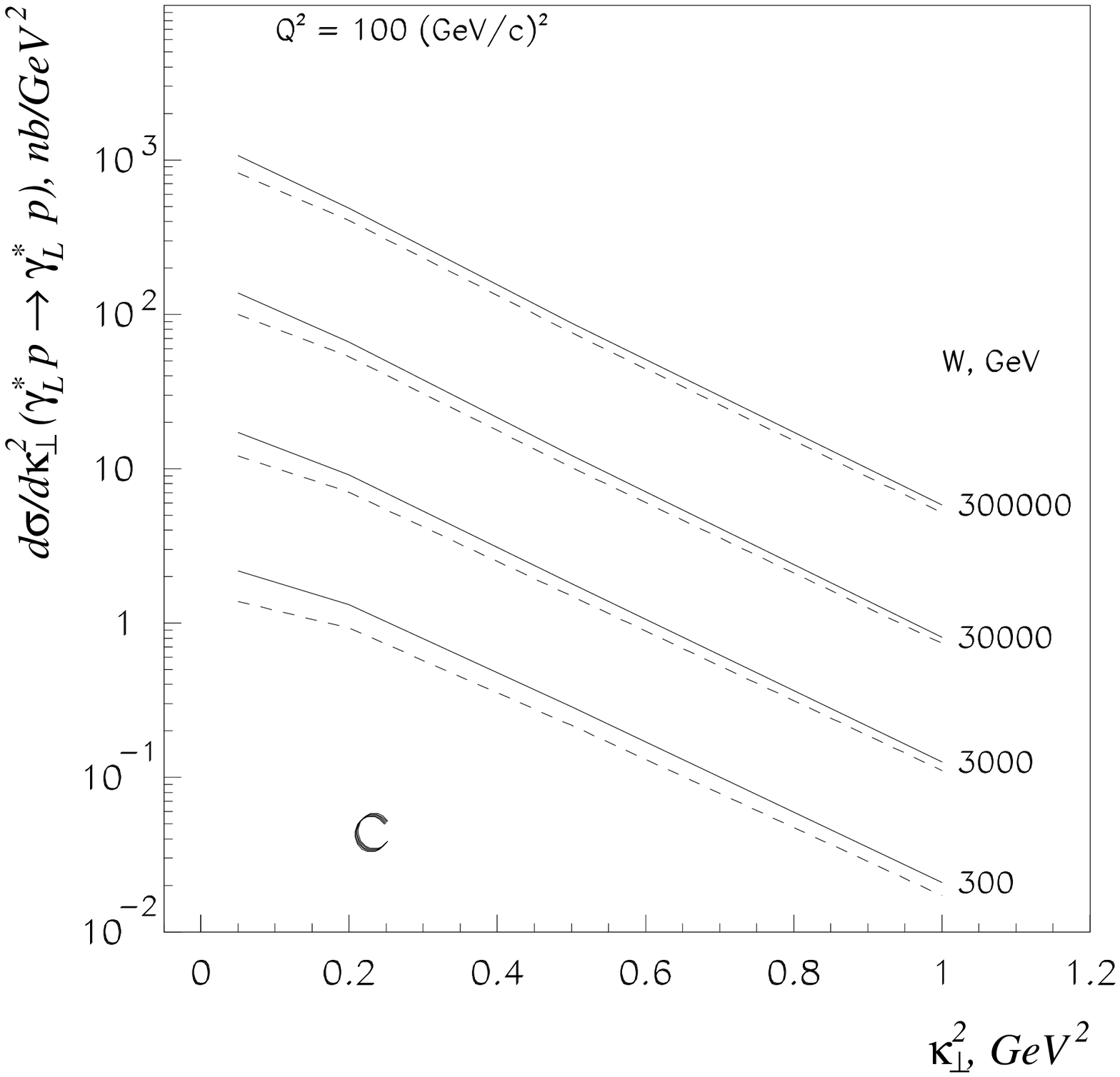,width=8.0cm}}
\vspace{1cm}
\caption{The same as in Fig. 4 for reaction
$\gamma^*_T(Q^2)p\to\gamma^*_T(Q^2)p$.}
\end{figure}

\begin{figure}
%fig.7
\centerline{
\epsfig{file=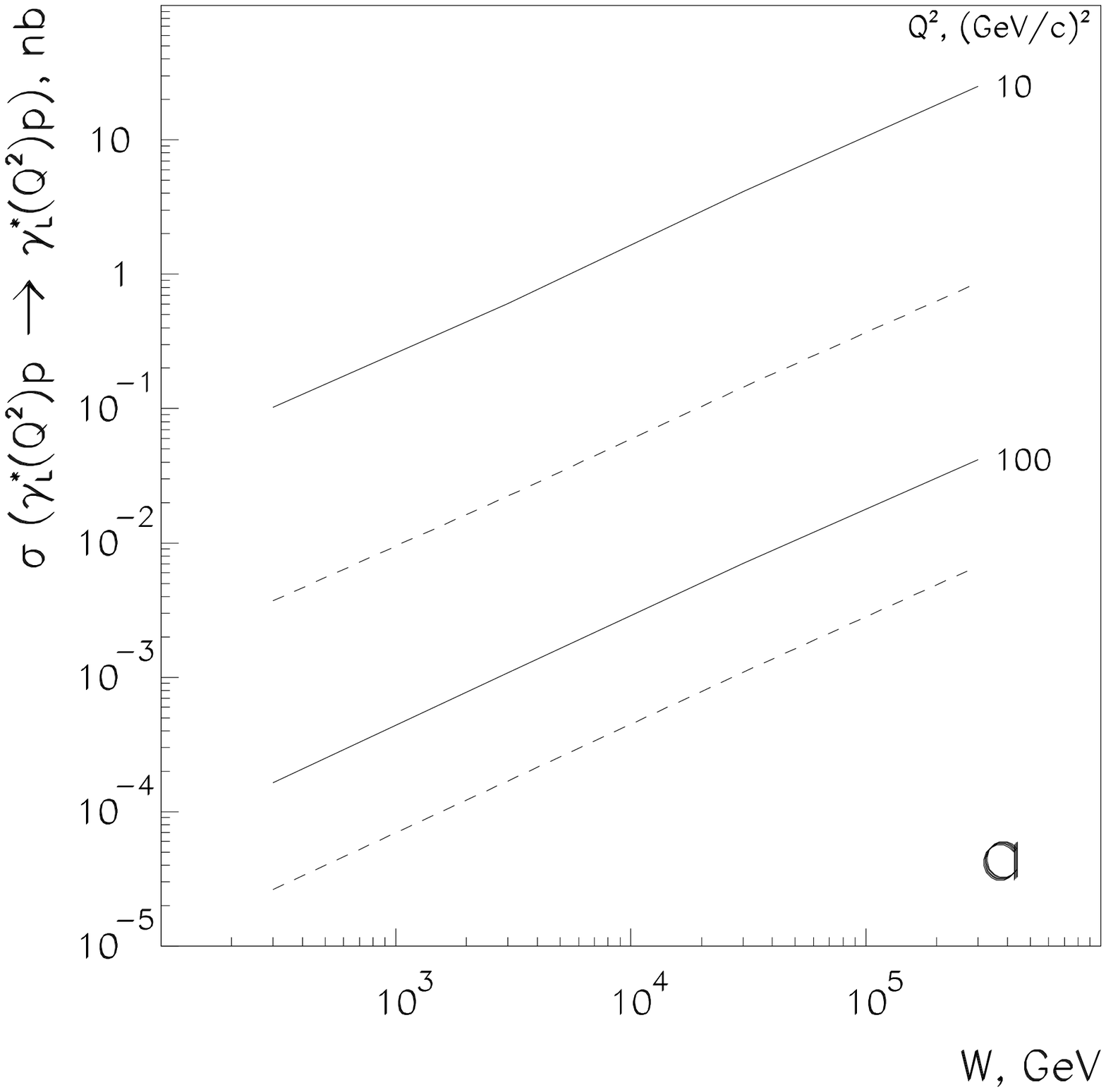,width=8.0cm}\hspace{1cm}
\epsfig{file=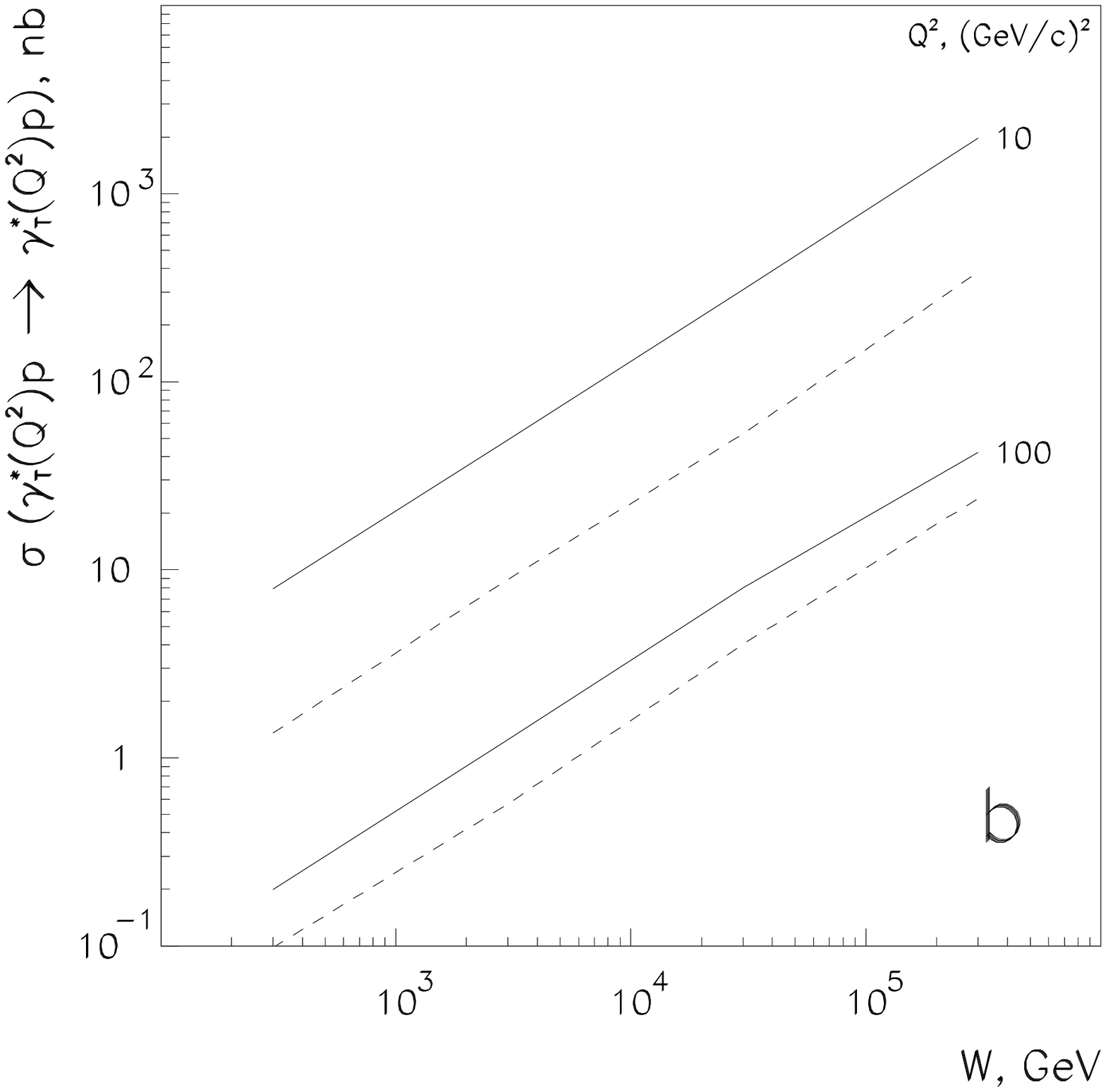,width=8.0cm}}
\vspace{1cm}
\caption{Cross sections
$\gamma^*_{L,T}(Q^2)p\to\gamma^*_{L,T}(Q^2)p$ with scale factor
$W^2/(M^2_{q\bar q}+M'^2_{q \bar q'})$.}
\end{figure}

\newpage
\begin{figure}
%fig.8
\centerline{
\epsfig{file=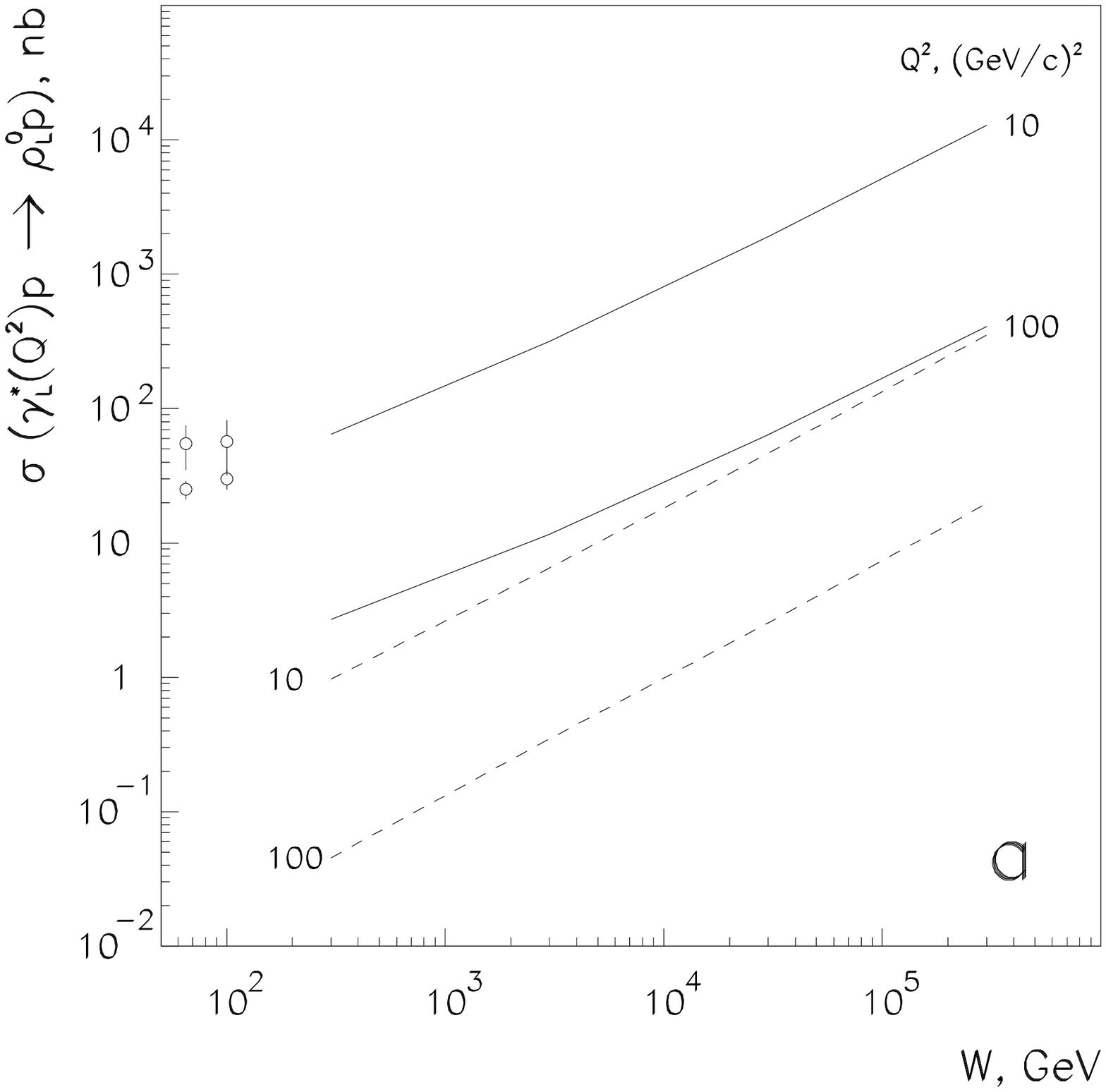,width=8.0cm}\hspace{1cm}
\epsfig{file=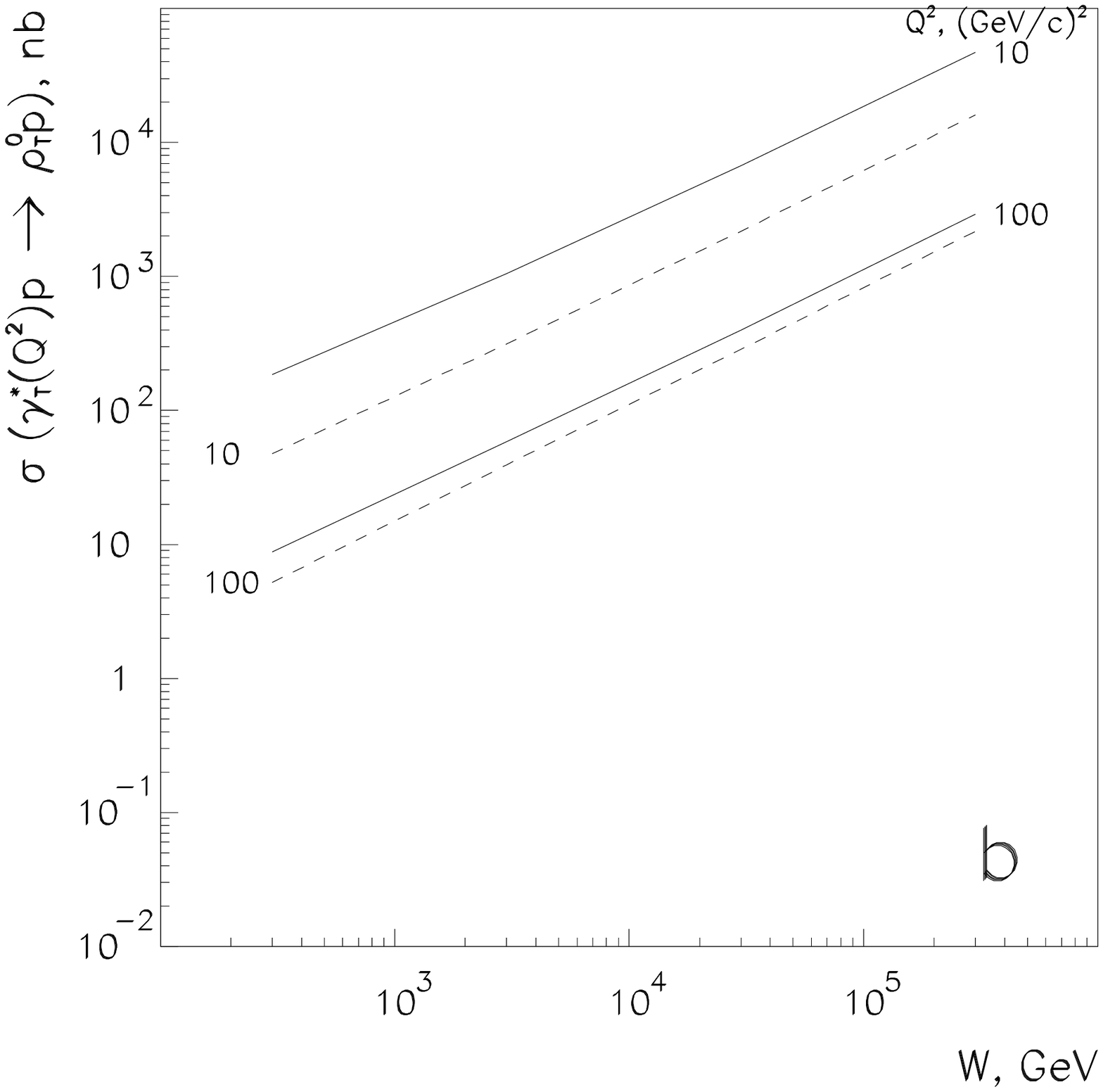,width=8.0cm}}
\vspace{1cm}
\centerline{
\epsfig{file=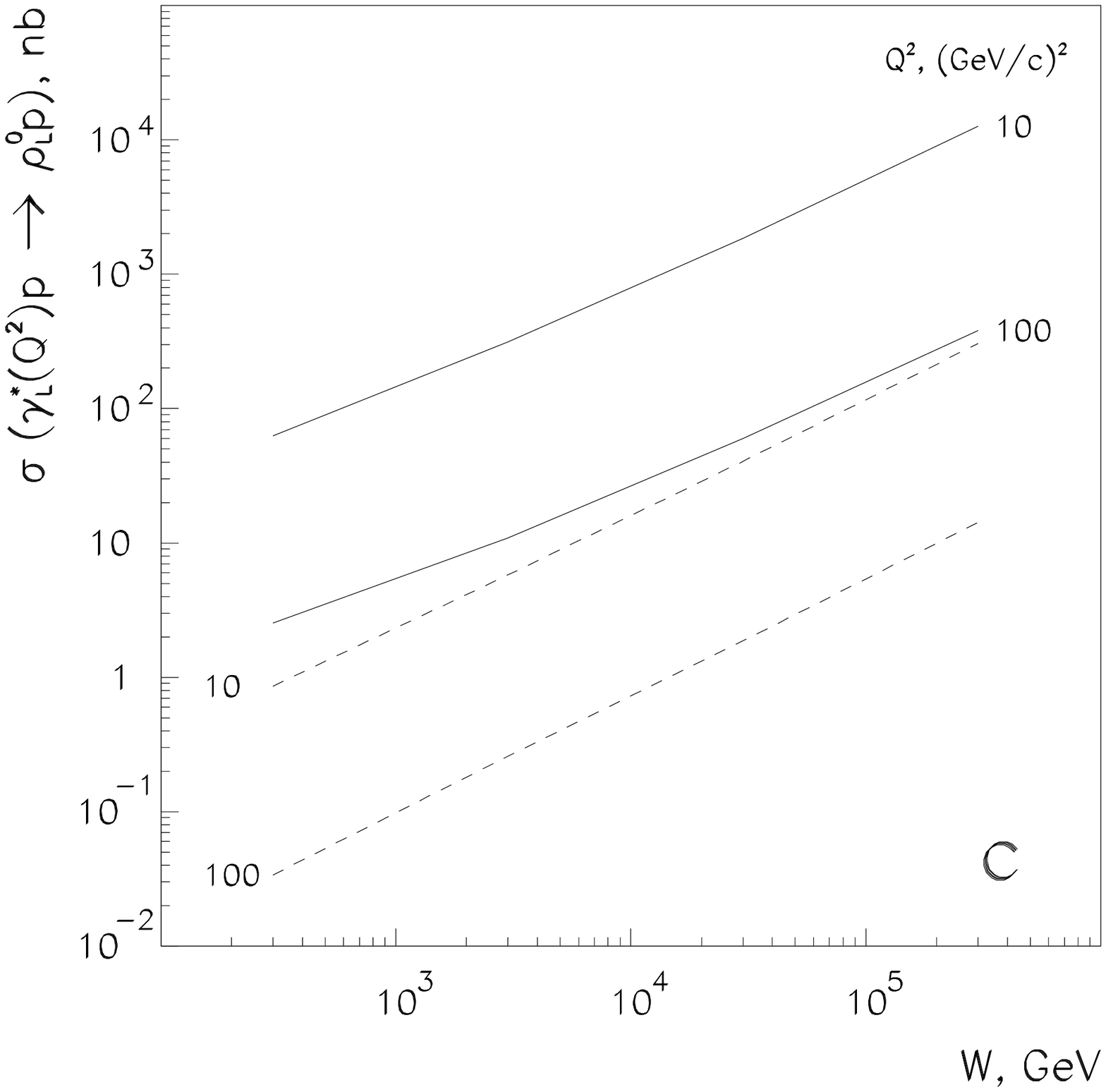,width=8.0cm}\hspace{1cm}
\epsfig{file=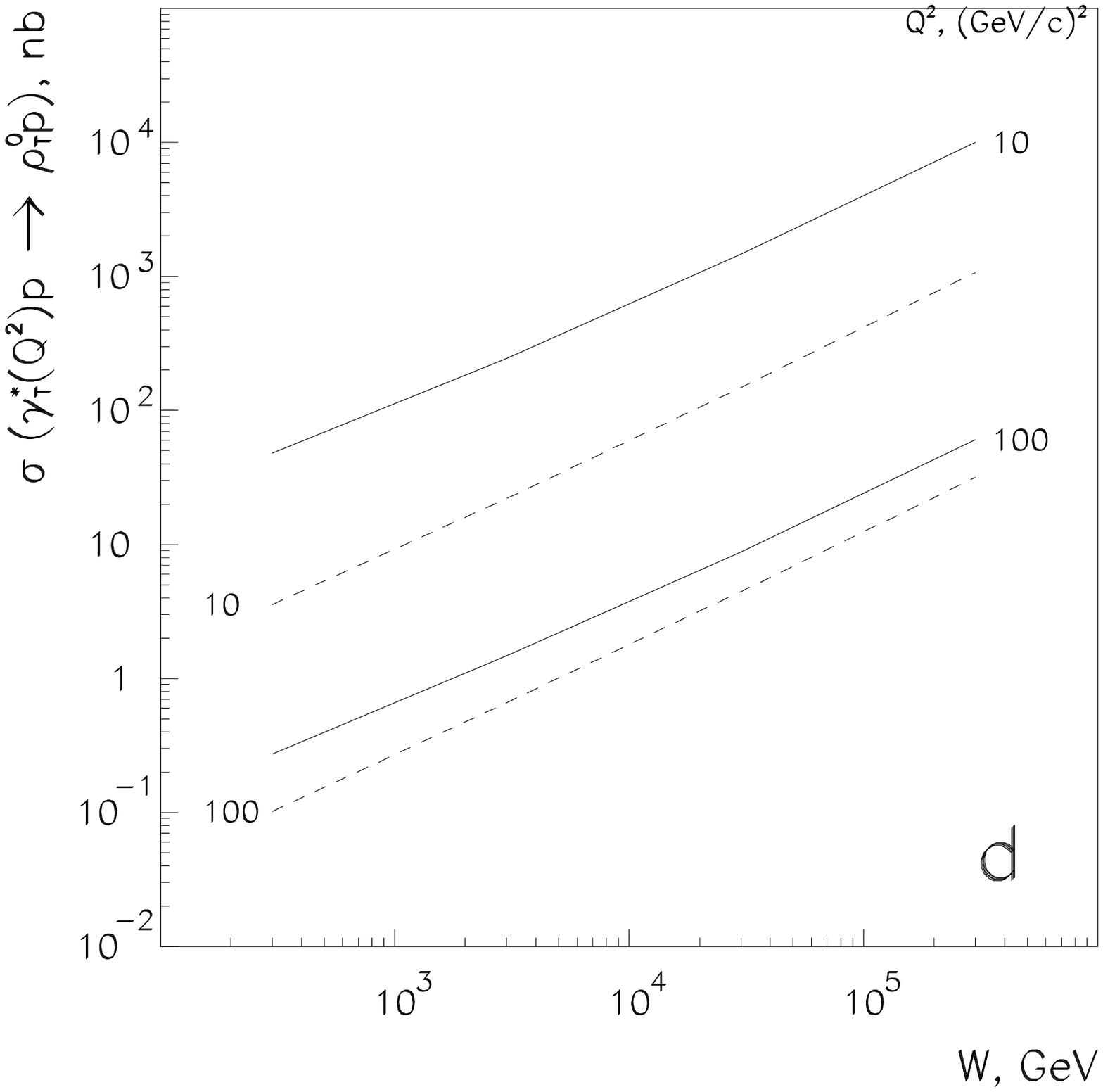,width=8.0cm}}
\vspace{1cm}
\caption{Cross sections in reactions
$\gamma^*_{L,T}(Q^2)p\to\rho^0_{L,T}p$ with scale factor
$W^2/(Q^2+\mu^2_V)$. a,b) Variant $G_\rho(M'^2_{q\bar q}) \sim
const$ at large $M'^2_{q\bar q}$; c,d)
$G_\rho(M'^2_{q\bar q})\sim 1/M'^2_{q\bar q}$ at large
$M'^2_{q\bar q}$. Experimental data shown in Fig. 8a are taken from
[18]; they are used for fixing the constant $C$ in Eq. (49).}
\end{figure}

\end{document}